\documentclass[12pt]{article}
\pdfoutput=1

\usepackage[a4paper,text={16.8cm,22.4cm}]{geometry}
\usepackage{amsmath,amsfonts,braket,slashed,amssymb,tikz,bm,psfrag,graphicx,color,dsfont,euscript}
\usepackage{mathtools}
\usepackage{multicol}
\usepackage{ctable}
\usepackage[small,labelfont=bf]{caption}

\RequirePackage[sort&compress,square,comma,numbers]{natbib}
\allowdisplaybreaks
\newcommand{\F}{{\EuScript F}}

\newcommand{\Ha}{{\EuScript H}}
\newcommand{\Sa}{{\EuScript S}}

\newcommand{\nsl}{\rlap{\hspace{0.25mm}/}{n}}
\newcommand{\nbsl}{\rlap{\hspace{0.25mm}/}{\bar n}}

\DeclarePairedDelimiterX\MeijerM[3]{\lparen}{\rparen}%
{\begin{smallmatrix}#1 \\[1mm] #2\end{smallmatrix}\delimsize\vert\,#3}

\newcommand\MeijerG[8][]{G^{\,#2,#3}_{#4,#5}\MeijerM[#1]{#6}{#7}{#8}}

\begin{document}

\begin{titlepage}

\begin{flushright}
\normalsize
MITP/20-023\\ 
May 6, 2020
\end{flushright}

\vspace{1.0cm}
\begin{center}
\Large\bf
Renormalization and Scale Evolution of the\\ Soft-Quark Soft Function
\end{center}

\vspace{0.5cm}
\begin{center}
Ze Long Liu$^{a}$, Bianka Mecaj$^b$, Matthias Neubert$^{b,c}$, Xing Wang$^b$ 
and Sean Fleming$^b$\footnote[0]{On leave from Department of Physics, University of Arizona, Tucson, AZ 85721, U.S.A.}\\
\vspace{0.7cm} 
{\sl ${}^a$Theoretical Division, Los Alamos National Laboratory, Los Alamos, NM 87545, U.S.A.\\[3mm]
${}^b$PRISMA$^+$ Cluster of Excellence \& Mainz Institute for Theoretical Physics\\
Johannes Gutenberg University, 55099 Mainz, Germany\\[3mm]
${}^c$Department of Physics \& LEPP, Cornell University, Ithaca, NY 14853, U.S.A.}
\end{center}

\vspace{0.8cm}
\begin{abstract}
Soft functions defined in terms of matrix elements of soft fields dressed by Wilson lines are central components of factorization theorems for cross sections and decay rates in collider and heavy-quark physics. While in many cases the relevant soft functions are defined in terms of gluon operators, at subleading order in power counting soft functions containing quark fields appear. We present a detailed discussion of the properties of the soft-quark soft function consisting of a quark propagator dressed by two finite-length Wilson lines connecting at one point. This function enters in the factorization theorem for the Higgs-boson decay amplitude of the $h\to\gamma\gamma$ process mediated by light-quark loops. We perform the renormalization of this soft function at one-loop order, derive its two-loop anomalous dimension and discuss solutions to its renormalization-group evolution equation in momentum space, in Laplace space and in the ``diagonal space'', where the evolution is strictly multiplicative. 
\end{abstract}

\end{titlepage}

\tableofcontents
\newpage

\section{Introduction}

Soft-collinear effective theory (SCET) offers a convenient framework for analyzing the factorization properties of cross sections and scattering amplitudes sensitive to different, hierarchical scales \cite{Bauer:2001yt,Bauer:2002nz,Beneke:2002ph,Becher:2014oda}. The corresponding factorization theorems contain hard functions, jet functions and soft functions, which receive contributions from different momentum regions in Feynman diagrams. The hard functions correspond to Wilson coefficients obtained when the full theory is matched onto SCET, while the jet and soft functions are defined in terms of matrix elements in the low-energy effective theory. Soft functions -- matrix elements of non-local products of soft fields dressed by Wilson lines -- play a particularly important role in the factorization theorems, because they often capture the physics at the longest relevant distance scales in a given process. In some cases the soft functions are non-perturbative objects, whereas in others they can be calculated using perturbation theory.

Recently, there has been a growing interest in understanding factorization at subleading power in scale ratios. In this case a large number of hard, jet and soft functions appear. In particular, while at leading power soft emissions are eikonal and can be described by soft Wilson lines, at subleading power the emission of soft fermions and power-suppressed emissions of soft gauge bosons need to be taken into account. Particularly interesting is the case of soft quark emission, which is absent at leading power. At subleading order in the SCET expansion there is a unique interaction that couples a soft quark to collinear quarks and gauge fields \cite{Beneke:2002ph}. 

As a concrete example, we have considered in \cite{Liu:2019oav} the case of the $h\to\gamma\gamma$ decay amplitude induced by $b$-quark loops. The relevant soft function is derived from the vacuum matrix element of the soft quark propagator dressed by two finite-length soft Wilson lines, i.e.\ 
\begin{equation}\label{softme}
   - \frac{(4\pi)^{1-\epsilon}}{N_c}\,e^{\epsilon\gamma_E}\,\mu^{2\epsilon}\,
   \langle 0|\,T\,\mbox{Tr}\,S_{\bar n}(0,r_1\bar n)\,q_s(r_1\bar n)\,
    \bar q_s(r_2 n)\,S_{n}(r_2 n,0)\,|0\rangle \,,
\end{equation}
where the trace is over color (but not spinor) indices, and $\epsilon=(4-d)/2$ is the dimensional regulator. The prefactor is chosen for later convenience. We have introduced two light-like reference vectors $n^\mu$ and $\bar n^\mu$ (with $n\cdot\bar n=2$) aligned with the directions of the two photons in the $h\to\gamma\gamma$ process. The soft quark fields are displaced from the Higgs vertex at position $z=0$ by light-like distances along the $n$ and $\bar n$ light cones. The Wilson lines $S_n$ and $S_{\bar n}$ connect the soft quarks with the Higgs vertex and ensure that the matrix element is gauge invariant. A different soft-quark soft function, which is relevant for inclusive cross sections, was introduced in \cite{Moult:2019mog} and has been studied further in \cite{Moult:2019uhz,Moult:2019vou}.

In the context of the SCET, soft quarks couple to collinear fields via subleading interactions in the SCET Lagrangian \cite{Beneke:2002ph}. In our case the soft matrix element in (\ref{softme}) is sandwiched between projection operators $P_n={\scriptsize \frac{\nsl\nbsl}{4}}$ from the right, where it connects to $n$-collinear fields, and $\bar P_{\bar n}={\scriptsize \frac{\nsl\nbsl}{4}}$ from the left, where it connects with $\bar n$-collinear fields. In the first step, we define a soft function ${\cal S}(\ell_+\ell_-)$ via the Fourier transform of the above matrix element with respect to the coordinates $r_1$ and $r_2$, 
\begin{equation}
\begin{aligned}
   \frac{i}{2}\,\,{\cal S}(\ell_+\ell_-)\,P_n
   &=  - \frac{(4\pi)^{1-\epsilon}}{N_c}\,e^{\epsilon\gamma_E}
    \int dr_1\,e^{ir_1\ell_-} \int dr_2\,e^{-ir_2\ell_+} \\
   &\quad\times \mu^{2\epsilon}\,
    \langle 0|\,T\,\mbox{Tr}\,\bar P_{\bar n}\,S_{\bar n}(0,r_1\bar n)\,q_s(r_1\bar n)\,
    \bar q_s(r_2 n)\,S_{n}(r_2 n,0)\,P_n\,|0\rangle \,.
\end{aligned}
\end{equation}
Reparameterization invariance \cite{Manohar:2002fd} ensures that the soft function ${\cal S}$ only depends on the product $w=\ell_+\ell_-$ of the Fourier variables $\ell_\pm$. At lowest order in perturbation theory (but not beyond) these variables can be identified with the light-cone components $n\cdot\ell$ and $\bar n\cdot\ell$ of the soft momentum $\ell^\mu$ flowing through the soft quark propagator. In the next step we define a new soft function $S(w)$ in terms of the discontinuity of the function ${\cal S}(\ell_+\ell_-)$, i.e.\
\begin{equation}\label{SDisc}
   S(w) = \frac{1}{2\pi i}\,\big[ {\cal S}(w+i0) - {\cal S}(w-i0) \big] \,.
\end{equation}
Our definition of the soft function differs from the one in \cite{Liu:2019oav} by a factor $(-N_c\alpha_b/\pi)$, where $\alpha_b=\alpha/9$ is the electromagnetic coupling of the $b$ quark.

In Section~\ref{sec:2} we collect the expression for the bare soft function obtained at one-loop order. A heuristic derivation of the non-local renormalization factor, which subtracts the divergences in the bare function, is presented in Section~\ref{sec:3} along with the expression for the renormalized soft function. In Section~\ref{sec:4} we discuss the renormalization-group (RG) evolution equation satisfied by the soft function and present a conjecture for its anomalous dimension at two-loop order. We also construct an exact solution to the RG equation at next-to-leading order (NLO) in RG-improved perturbation theory. The asymptotic behavior of the RG-improved soft function for large values $w\gg m_b^2$ is studied in Section~\ref{sec:6}, where we introduce the concept of dynamical scale setting. Section~\ref{sec:Laplace} contains a discussion of the properties of the Laplace transform of the soft function, which satisfies a novel form of RG evolution equation. In Section~\ref{sec:dual} we study the soft function in the ``diagonal space'', in which its RG evolution is strictly local in $w$. To this end we generalize the concept of the ``dual space'' introduced in \cite{Bell:2013tfa,Braun:2014owa} to higher orders of perturbation theory. Finally, in Section~\ref{sec:7} we analyze in detail the double convolution integral $T_3$ arising in the factorization theorem for the $h\to\gamma\gamma$ decay amplitude \cite{Liu:2019oav} and show that, after a suitable rapidity regularization in the diagonal space, this quantity is RG invariant and free of endpoint divergences. Section~\ref{sec:8} contains a summary of our main results and some conclusions. Several technical details of our analysis are presented in three appendices.

\section{One-loop expression for the bare soft function}
\label{sec:2}

\begin{figure}
\begin{center}
\includegraphics[width=0.7\textwidth]{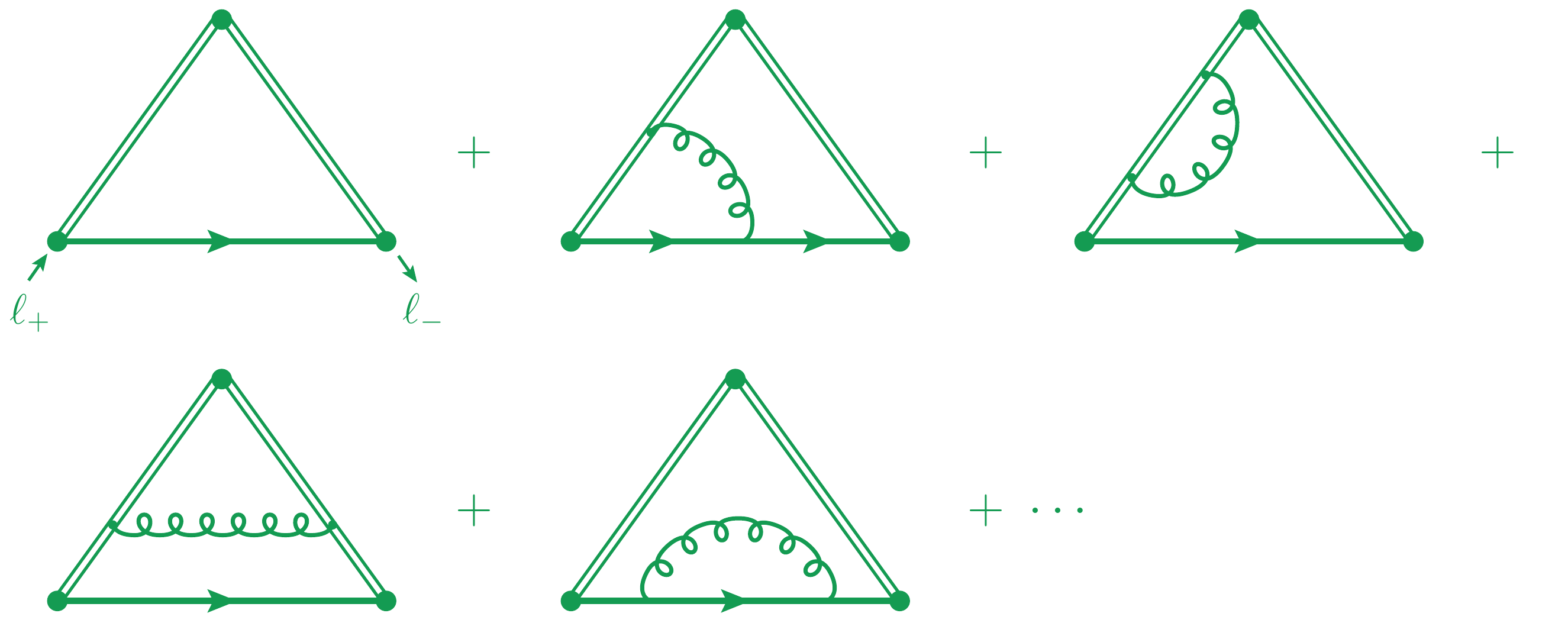}
\vspace{2mm}
\caption{\label{fig:Sgraphs} 
Representative Feynman diagrams contributing to the soft function ${\cal S}(\ell_+\ell_-)$ up to ${\cal O}(\alpha_s)$ (taken from \cite{Liu:2019oav}). We use a double-line notation to represent the finite-length soft Wilson lines. The three points mark the vertices connecting to the incoming Higgs boson (top) and the outgoing photons (bottom). The soft function $S(w)$ in (\ref{SDisc}) is given by the discontinuity of ${\cal S}(w)$.} 
\end{center}
\end{figure}

At one-loop order in perturbation theory the bare soft function has been calculated to all orders in the dimensional regulator $\epsilon=(4-d)/2$ in terms of hypergeometric functions \cite{Liu:2019oav}. The relevant Feynman graphs are shown in Figure~\ref{fig:Sgraphs}. Due to the multipole expansion applied to soft fields in interaction terms with collinear fields, the soft momentum component $\ell_+$ enters at the left lower vertex, while $\ell_-$ goes out at the right lower vertex, as indicated in the first graph. In more complicated diagrams, such as the second graph, the assignment of momenta becomes non-trivial due to the presence of the Wilson lines (see \cite{Liu:2019oav} for a detailed discussion). As a result, one finds that the soft function $S(w)$ defined via the discontinuity of the diagrams shown in Figure~\ref{fig:Sgraphs} has support for all values $w>0$, even though at leading order (first graph) the discontinuity arises only if $w>m_b^2$. When expanded about $\epsilon=0$ the result reads 
\begin{equation}
   S^{(0)}(w) = m_{b,0}\,\mu^{2\epsilon}\,
    \Big[ S_a^{(0)}(w)\,\theta(w-m_{b,0}^2) + S_b^{(0)}(w)\,\theta(m_{b,0}^2-w) \Big] \,,
\end{equation}
where
\begin{equation}\label{SaSb}
\begin{aligned}
   S_a^{(0)}(w) &= \frac{e^{\epsilon\gamma_E}}{\Gamma(1-\epsilon)}\,\big(w-m_{b,0}^2\big)^{-\epsilon} 
    \left[ 1 + \epsilon\,\frac{C_F\alpha_{s,0}}{4\pi}\,2e^{\epsilon\gamma_E}\,
    \frac{3-2\epsilon}{1-2\epsilon}\,\Gamma(\epsilon)\,
    \frac{\left(m_{b,0}^2\right)^{1-\epsilon}}{w-m_{b,0}^2} \right] \\
   &\quad\mbox{}+ \frac{C_F\alpha_{s,0}}{4\pi} \bigg[ 
    \left( - \frac{2}{\epsilon^2} + \frac{6}{\epsilon} 
    + \frac{2}{\epsilon}\,\ln\!\Big(1-\frac{1}{\hat w_0}\Big) + 12 - \frac{\pi^2}{3} \right) 
    \big(w-m_{b,0}^2\big)^{-2\epsilon} \\
   &\hspace{2.45cm}\mbox{}- 2\,\text{Li}_2\Big(\frac{1}{\hat w_0}\Big)
    - 2 \left( \ln \hat w_0 - 1 \right) \ln\!\Big(1-\frac{1}{\hat w_0}\Big) 
    - 3 \ln^2\!\Big(1-\frac{1}{\hat w_0}\Big) + {\cal O}(\epsilon) \bigg] \,, \\
   S_b^{(0)}(w) &= \frac{C_F\alpha_{s,0}}{4\pi}\,\big( m_{b,0}^2 \big)^{-2\epsilon}\,\bigg[ 
    - \frac{4}{\epsilon}\,\ln(1-\hat w_0) + 6 \ln^2(1-\hat w_0) + {\cal O}(\epsilon) \bigg] \,.
\end{aligned}
\end{equation}
Here $\alpha_{s,0}$ is the bare QCD coupling and $m_{b,0}$ denotes the bare mass of the $b$-quark. We have pulled out a factor $e^{\epsilon\gamma_E}/\left(4\pi\right)^\epsilon$ from the bare coupling, as appropriate in the $\overline{\rm MS}$ scheme. Moreover, we have defined the dimensionless ratio $\hat w_0=w/m_{b,0}^2$. 

To express the result in terms of physical parameters we renormalize the coupling according to $\alpha_{s,0}=\mu^{2\epsilon}\,Z_\alpha\,\alpha_s(\mu)$, where $Z_\alpha=1+{\cal O}(\alpha_s)$. From now on $\alpha_s\equiv\alpha_s(\mu)$ always denotes the renormalized coupling. The most convenient scheme for the renormalization of the $b$-quark mass is the pole scheme, in which the mass is defined by the position of the pole in the renormalized quark propagator \cite{Tarrach:1980up}. We denote the pole mass by $m_b$. At one-loop order one finds that
\begin{equation}
   \delta m_b^2 = m_b^2 - m_{b,0}^2 
   = \frac{C_F\alpha_{s,0}}{4\pi}\,2e^{\epsilon\gamma_E}\,\frac{3-2\epsilon}{1-2\epsilon}\,
    \Gamma(\epsilon) \left(m_{b,0}^2\right)^{1-\epsilon} \,.
\end{equation}
Renormalizing the quark mass in the pole scheme removes entirely the loop correction to the lowest-order term shown in the first line of the expression for $S_a(w)$ in (\ref{SaSb}). After the bare parameters have been renormalized, we obtain
\begin{equation}\label{Sstep1}
   S^{(0)}(w) = m_b\,\Big[ S_a^{(0)}(w)\,\theta(w-m_b^2) + S_b^{(0)}(w)\,\theta(m_b^2-w) \Big] \,,
\end{equation}
where 
\begin{equation}\label{SaSbren}
\begin{aligned}
   S_a^{(0)}(w) &= \left[ 1 + \frac{C_F\alpha_s}{4\pi} \left( - \frac{3}{\epsilon}
    + 3\ln\frac{m_b^2}{\mu^2} - 4 + {\cal O}(\epsilon) \right) \right]
    \frac{e^{\epsilon\gamma_E}}{\Gamma(1-\epsilon)}\,
    \bigg( \frac{w-m_b^2}{\mu^2} \bigg)^{-\epsilon} \\
   &\quad\mbox{}+ \frac{C_F\alpha_s}{4\pi} \bigg[ \left( - \frac{2}{\epsilon^2} + \frac{6}{\epsilon} 
    + \frac{2}{\epsilon}\,\ln\!\Big(1-\frac{1}{\hat w}\Big) + 12 - \frac{\pi^2}{3} \right) 
    \bigg( \frac{w-m_b^2}{\mu^2} \bigg)^{-2\epsilon} \\
   &\hspace{2.25cm}\mbox{}- 2\,\text{Li}_2\Big(\frac{1}{\hat w}\Big)
    - 2 \left( \ln \hat w - 1 \right) \ln\!\Big(1-\frac{1}{\hat w}\Big) 
    - 3 \ln^2\!\Big(1-\frac{1}{\hat w}\Big) + {\cal O}(\epsilon) \bigg] \,, \\
   S_b^{(0)}(w) &= \frac{C_F\alpha_s}{4\pi}\,\bigg( \frac{m_b^2}{\mu^2} \bigg)^{-2\epsilon}\,\bigg[ 
    - \frac{4}{\epsilon}\,\ln(1-\hat w) + 6 \ln^2(1-\hat w) + {\cal O}(\epsilon) \bigg] \,,
\end{aligned}
\end{equation}
where now $\hat w=w/m_b^2$. The remaining $1/\epsilon^n$ poles must be removed by renormalizing the soft function itself. A feature that is at first sight surprising is the appearance of the $1/\epsilon$ pole in the function $S_b^{(0)}$, which vanishes at lowest order. It is obvious that this pole can only be removed by means of a renormalization factor $Z_S(w,w';\mu)$ that is non-local in the space of $w$ values. In the following section we will present a conjecture for this renormalization factor and show that it indeed removes all remaining singularities.

\section{Renormalization of the soft function}
\label{sec:3}

The soft-quark soft function appears in the analysis of the factorization theorem for the amplitude for the radiative Higgs-boson decay $h\to\gamma\gamma$ induced by loops containing light $b$-quarks \cite{Liu:2019oav}. While this is not the dominant contribution to the decay amplitude (which instead comes from loops containing top-quarks or $W$ bosons), this particular contribution has an interesting structure, since it receives large double-logarithmic corrections of order $\alpha\alpha_s^n L^{2n+2}$ with $n\in\mathbb{N}_0$, where $L=\ln(M_h^2/m_b^2)-i\pi$ \cite{Kotsky:1997rq,Akhoury:2001mz,Penin:2014msa,Liu:2017vkm,Liu:2018czl}. $M_h\gg m_b$ denotes the mass of the Higgs boson. In a recent paper \cite{Liu:2019oav}, two of us have derived a factorization theorem in SCET, with which these large logarithms can eventually be resummed to all orders in perturbation theory. The factorization theorem contains three terms: a term $T_1$ involving a hard function $H_1$ multiplied with the $h\to\gamma\gamma$ matrix element of a local SCET operator containing a Higgs field and two photon fields, a second term $T_2$ involving the convolution of a hard function $H_2$ with a non-local SCET operator containing a Higgs field, a photon field and two collinear quark fields, and a third term $T_3$ containing a hard function $H_3$ multiplied with a double convolution of the soft-quark soft function with two jet functions. In the first two terms the SCET matrix elements live at the scale $m_b$, however the matrix element in third term depends in addition on an intermediate scale of order $\sqrt{M_h m_b}$. We have presented arguments in \cite{Liu:2019oav} suggesting that the third term should by itself be RG invariant. More accurately, we found that when this term is evaluated as a convolution over bare functions, with cutoffs implemented to remove endpoint divergences, almost all $1/\epsilon^n$ poles cancel out. A single remaining pole term proportional to $\zeta_3/\epsilon$ can be attributed to the fact that the implementation of cutoffs on the bare functions is not strictly compatible with RG invariance (see also the discussion in Section~\ref{sec:7}).

We will now rely on these arguments to derive the renormalization condition for the soft function. 
In terms of bare functions, the third term of the factorization theorem for the $b$-quark induced $h\to\gamma\gamma$ decay amplitude takes the form (for now we omit the cutoffs required to defined this expression properly)
\begin{equation}\label{T3}
   T_3 = H_3^{(0)} \int_0^\infty\!\frac{d\ell_-}{\ell_-} \int_0^\infty\!\frac{d\ell_+}{\ell_+}\,
    J^{(0)}(M_h\ell_-)\,J^{(0)}(-M_h\ell_+)\,S^{(0)}(\ell_+\ell_-) \,.
\end{equation}
Note that one of the jet functions $J^{(0)}(p^2)$ is evaluated at time-like momentum ($p^2>0$) and the other one at space-like momentum ($p^2<0$). The renormalization of the hard function is well understood, with the corresponding anomalous dimension being known to three-loop order \cite{Becher:2006mr}. The renormalization of the jet function $J(p^2)$ was derived at one-loop order a long time ago \cite{Bosch:2003fc} and has recently been extended to two loops \cite{Liu:2020ydl}. One finds that the time-like (space-like) jet functions at different $p^2$ values mix with each other. The renormalization conditions can be written in the form 
\begin{equation}\label{HJren}
\begin{aligned}
   H_3(\mu) &= Z_{33}^{-1}(\mu)\,H_3^{(0)} \,, \\
   J(\pm M_h\ell,\mu) &= \int_0^\infty\!d\ell'\,Z_J(\pm M_h\ell,\pm M_h\ell';\mu)\,
    J^{(0)}(\pm M_h\ell') \,,
\end{aligned}   
\end{equation}
where at one-loop order
\begin{equation}\label{Zfactors}
\begin{aligned}
   Z_{33}^{-1}(\mu) &= 1 + \frac{C_F\alpha_s}{4\pi} \left[ \frac{2}{\epsilon^2} 
    - \frac{2}{\epsilon} \left( \ln\frac{-M_h^2}{\mu^2} - \frac32 \right) \right] 
    + {\cal O}(\alpha_s^2) \,, \\
   Z_J(\pm M_h\ell,\pm M_h\ell';\mu) 
   &= \left[ 1 + \frac{C_F\alpha_s}{4\pi} \left( - \frac{2}{\epsilon^2} 
    + \frac{2}{\epsilon} \ln\frac{\mp M_h\ell}{\mu^2} \right) \right] \delta(\ell-\ell') + \frac{C_F\alpha_s}{2\pi\epsilon}\,\ell\,\Gamma(\ell,\ell')
    + {\cal O}(\alpha_s^2) \,.
\end{aligned}
\end{equation}
Here and below $-M_h^2\equiv-M_h^2-i0$ and $-p^2\equiv-p^2-i0$. We have introduced the symmetric distribution \cite{Bosch:2003fc}
\begin{equation}\label{Gammadef}
   \Gamma(\omega,\omega') = \left[ \frac{\theta(\omega-\omega')}{\omega(\omega-\omega')}
    + \frac{\theta(\omega'-\omega)}{\omega'(\omega'-\omega)} \right]_+ .
\end{equation}
The plus prescription is defined such that, when $\Gamma(\omega,\omega')$ is integrated with a function $f(\omega')$, one must replace $f(\omega')\to f(\omega')-f(\omega)$ under the integral. 

Using the inverse of the relations (\ref{HJren}) we can express the first three functions in the double convolution for $T_3$ in (\ref{T3}) in terms of their renormalized counterparts. Requiring that the combined expression for $T_3$ is scale independent we then obtain
\begin{equation}\label{T3ren}
   T_3 = H_3(\mu) \int_0^\infty\!\frac{d\ell_-}{\ell_-}
    \int_0^\infty\!\frac{d\ell_+}{\ell_+}\,J(M_h\ell_-,\mu)\,J(-M_h\ell_+,\mu)\,
    S(\ell_+\ell_-,\mu) \,,
\end{equation}
where the renormalized soft function is defined as
\begin{equation}
\begin{aligned}
   S(\ell_+\ell_-,\mu) 
   &= Z_{33}(\mu)\,\int_0^\infty\!d\ell_-'\,\frac{\ell_-}{\ell_-'}\,
    Z_J^{-1}(M_h\ell_-',M_h\ell_-;\mu) \\
   &\hspace{1.3cm}\times \int_0^\infty\!d\ell_+'\,\frac{\ell_+}{\ell_+'}\,
    Z_J^{-1}(-M_h\ell_+',-M_h\ell_+;\mu)\,S^{(0)}(\ell_+'\ell_-') \,.
\end{aligned}
\end{equation}
At one-loop order the inverse renormalization factors are given by the same expressions as in (\ref{Zfactors}), but with the signs in front of $\alpha_s$ reversed. The above renormalization condition can be simplified by carefully evaluating the convolution of the three $Z$ factors. After some algebra we find the simple form 
\begin{equation}\label{RGES}
   S(w,\mu) = \int_0^\infty\!dw'\,Z_S(w,w';\mu)\,S^{(0)}(w') \,,
\end{equation}
where
\begin{equation}\label{ZS1loop}
   Z_S(w,w';\mu) = \left[ 1 + \frac{C_F\alpha_s}{4\pi} \left( \frac{2}{\epsilon^2}
    - \frac{2}{\epsilon} \ln\frac{w}{\mu^2} - \frac{3}{\epsilon} \right) \right] \delta(w-w') 
    - \frac{C_F\alpha_s}{\pi\epsilon}\,w\,\Gamma(w,w') + {\cal O}(\alpha_s^2) \,. 
\end{equation}
Note that the same plus distribution $\Gamma(w,w')$ as in (\ref{Gammadef}) appears, and that the logarithmic terms contained in $Z_{33}$ and $Z_J^{-1}$ have conspired to generate a logarithm of the ratio $w/\mu^2$. This fact has already been anticipated in \cite{Liu:2019oav}.

\begin{figure}
\begin{center}
\includegraphics[width=0.6\textwidth]{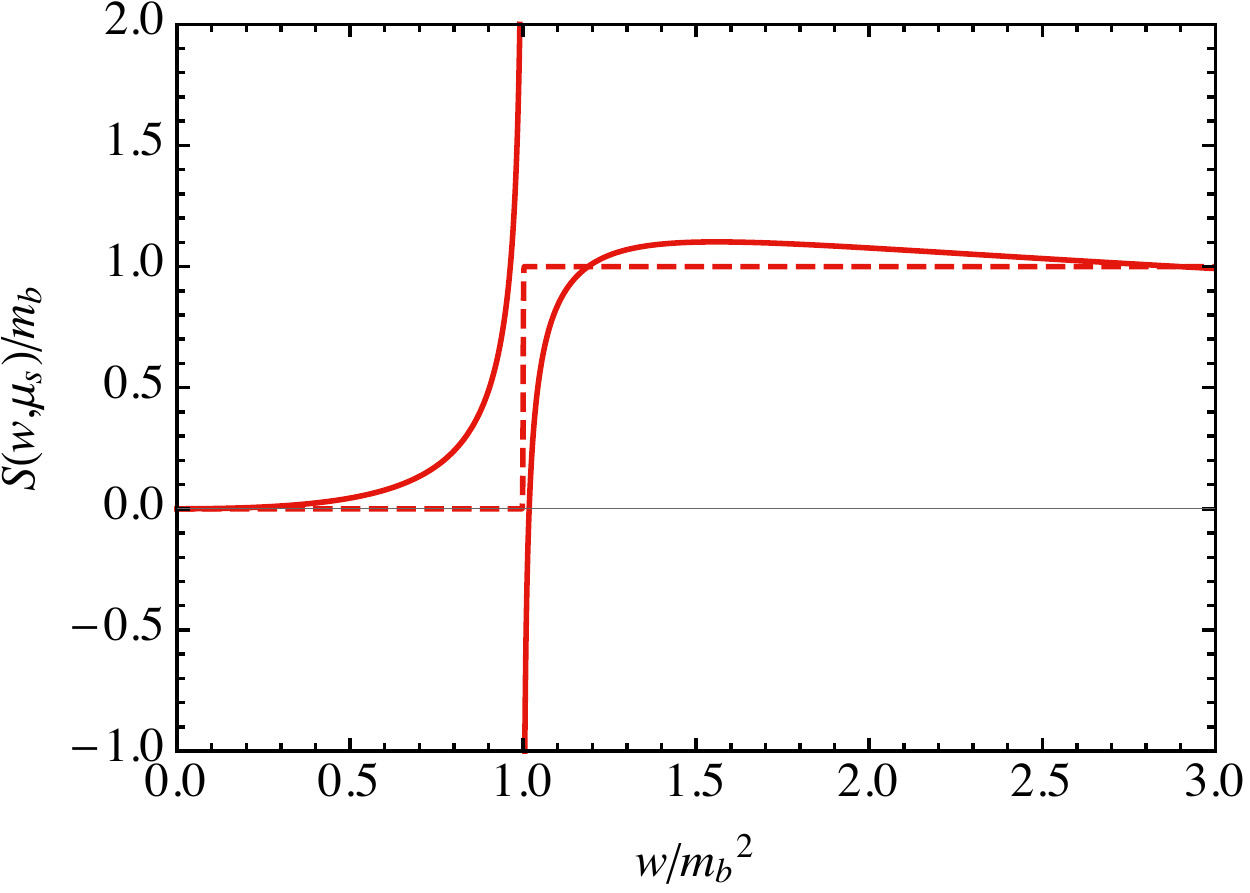}
\vspace{2mm}
\caption{\label{fig:S1loop} 
Renormalized soft function $S(w,\mu)/m_b$ for $\mu=m_b$ at tree level (dashed) and one-loop order (solid).} 
\end{center}
\end{figure}

It is not at all obvious that this definition of the renormalized soft function ensures that all $1/\epsilon^n$ pole terms in (\ref{SaSbren}) are removed. Our ``derivation'' of the renormalization factor $Z_S$ is only a conjecture, since the convolution of the four bare functions in (\ref{T3}) contains endpoint divergences for $\ell_\pm\to\infty$ and is thus ill-defined. Nevertheless, applying the renormalization condition (\ref{RGES}) to the bare soft function in (\ref{Sstep1}) we find that the renormalization factor $Z_S(w,w';\mu)$ indeed removes all $1/\epsilon^n$ pole terms. Note that (contrary to the one-loop renormalization of the jet function \cite{Bosch:2003fc}) the plus distributions have a non-trivial effect, because the lowest-order soft function is not constant, see (\ref{SaSbren}). For the renormalized soft function at one-loop order we obtain
\begin{equation}\label{softrenorm}
   S(w,\mu) = m_b\,\Big[ S_a(w,\mu)\,\theta(w-m_b^2) + S_b(w,\mu)\,\theta(m_b^2-w) \Big] \,,
\end{equation}
with
\begin{equation}\label{eqs18}
\begin{aligned}
   S_a(w,\mu) &= 1 + \frac{C_F\alpha_s}{4\pi} \bigg[ - L_w^2 - 6 L_w + 3 L_m + 8 - \frac{\pi^2}{2} 
    + 2 \mbox{Li}_2\Big(\frac{1}{\hat w}\Big) \\
   &\hspace{2.53cm}\mbox{}- 4\ln\!\Big(1-\frac{1}{\hat w}\Big)\,
    \bigg( L_m + 1 + \ln\!\Big(1-\frac{1}{\hat w}\Big) + \frac32 \ln\hat w \bigg) \bigg] \,, \\
   S_b(w,\mu) &= \frac{C_F\alpha_s}{\pi}\,\ln(1-\hat w)\,\big[ L_m + \ln(1-\hat w) \big] \,.
\end{aligned}
\end{equation}
We have defined $L_w=\ln(w/\mu^2)$ and $L_m=\ln(m_b^2/\mu^2)$. Figure~\ref{fig:S1loop} shows the renormalized soft function $S(w,\mu)$ in units of $m_b$ as a function of the dimensionless ratio $\hat w=w/m_b^2$. We use $m_b=4.8$\,GeV for the $b$-quark pole mass and choose $\mu=m_b$ for the renormalization scale. Note that both curves are discontinuous at $\hat w=1$. 

\section{Renormalization-group evolution}
\label{sec:4}

The dependence of the renormalized soft function on the scale $\mu$ can be controlled by means of the RG evolution equation
\begin{equation}\label{Sevol}
   \frac{d}{d\ln\mu}\,S(w,\mu) = - \int_0^\infty\!dw'\,\gamma_S(w,w';\mu)\,S(w',\mu) \,,
\end{equation}
where the anomalous dimension is obtained from the coefficient of the single $1/\epsilon$ pole in $Z_S$ via 
\begin{equation}\label{eq22}
\begin{aligned}
   \gamma_S(w,w';\mu) 
   &= 2\alpha_s\,\frac{\partial Z_S^{[1]}(w,w';\mu)}{\partial\alpha_s} \\
   &= - \frac{C_F\alpha_s}{4\pi} \left[  \left( 4\ln\frac{w}{\mu^2} + 6 \right) \delta(w-w') 
    + 8 w\,\Gamma(w,w') \right] + {\cal O}(\alpha_s^2) \,. 
\end{aligned}
\end{equation}
We have checked explicitly that our expression for the renormalized soft function in (\ref{softrenorm}) satisfies the evolution equation (\ref{Sevol}) at ${\cal O}(\alpha_s)$. 

\subsection{Two-loop anomalous dimension}

We can derive additional information by using the known RG equations obeyed by the hard and jet functions. To all orders in perturbation theory, the evolution equation for the hard function takes the form \cite{Becher:2006mr,Becher:2009qa}
\begin{equation}\label{RGEH3}
   \frac{d}{d\ln\mu}\,H_3(\mu) 
   = \left[ \Gamma_{\rm cusp}(\alpha_s)\,\ln\frac{-M_h^2}{\mu^2} + 2\gamma_q(\alpha_s) \right]
    H_3(\mu) \,,
\end{equation}
where $\Gamma_{\rm cusp}$ is the light-like cusp anomalous dimension in the fundamental representation of SU$(N_c)$ \cite{Korchemskaya:1992je}, and $\gamma_q$ is the anomalous dimension of the quark field in light-cone gauge \cite{Becher:2009qa}. The cusp anomalous dimension is known to four-loop order while $\gamma_q$ is known to three loops. The anomalous dimension for the jet function has only recently been calculated at two-loop order \cite{Liu:2020ydl}. The renormalized jet functions in (\ref{T3ren}) obey the RG evolution equation
\begin{equation}\label{RGE}
   \frac{d}{d\ln\mu}\,J(p^2,\mu) 
   = - \int_0^\infty\!dx\,\gamma_J(p^2,x p^2;\mu)\,J(x p^2,\mu) \,.
\end{equation}
Compared with (\ref{Zfactors}) we have introduced a dimensionless integration variable $x$, such that this relation holds for both time-like and space-like values of $p^2$. The anomalous dimension is given by
\begin{equation}\label{gammaJ}
\begin{aligned}
   \gamma_J(p^2,x p^2;\mu) 
   &= \left[ \Gamma_{\rm cusp}(\alpha_s)\,\ln\frac{-p^2}{\mu^2}
    - \gamma'(\alpha_s) \right] \delta(1-x) + \Gamma_{\rm cusp}(\alpha_s)\,\Gamma(1,x) \\
   &\quad\mbox{}+ C_F \left( \frac{\alpha_s}{2\pi} \right)^2 \frac{\theta(1-x)}{1-x}\,h(x) 
    + {\cal O}(\alpha_s^3) \,,
\end{aligned}
\end{equation}
where 
\begin{equation}\label{hdef}
   h(x) = \ln x \left[ \beta_0 
    + 2C_F \left( \ln x - \frac{1+x}{x}\,\ln(1-x) - \frac32 \right) \right] .
\end{equation}
The local terms (with $x=1$) can to all orders by expressed in terms of the cusp anomalous dimension and an anomalous dimension $\gamma'(\alpha_s)$. Since the plus distribution contained in $\Gamma(1,x)$ is linked with the logarithmic term, it is also multiplied by $\Gamma_{\rm cusp}$. However, starting at two-loop order additional non-local terms arise, whose explicit form was obtained in \cite{Liu:2020ydl} by using the RG invariance of the $B^-\to\gamma\ell^-\bar\nu$ decay rate along with the calculation of the two-loop anomalous dimension of the $B$-meson light-cone distribution amplitude (LCDA) performed in~\cite{Braun:2019wyx}.

With the help of the above expressions we can write the anomalous dimension for the soft function defined in (\ref{Sevol}) in the more general form
\begin{equation}\label{gammaS}
\begin{aligned}
   \gamma_S(w,w';\mu) 
   &= - \left[ \Gamma_{\rm cusp}(\alpha_s)\,\ln\frac{w}{\mu^2} - \gamma_s(\alpha_s) \right]
    \delta(w-w') - 2\Gamma_{\rm cusp}(\alpha_s)\,w\,\Gamma(w,w') \\
   &\quad\mbox{}- 2 C_F \left( \frac{\alpha_s}{2\pi} \right)^2 
    \frac{w\,\theta(w'-w)}{w'(w'-w)}\,h\bigg(\frac{w}{w'}\bigg) + {\cal O}(\alpha_s^3) \,,
\end{aligned}
\end{equation}
where 
\begin{equation}\label{gams}
   \gamma_s(\alpha_s) = 2\gamma_q(\alpha_s) + 2\gamma'(\alpha_s) \,.
\end{equation}
Explicit expressions for the various anomalous dimensions are given in Appendix~\ref{app:A}. While the one-loop expression for the soft anomalous dimension shown in (\ref{eq22}) is derived from the one-loop renormalization factor $Z_S$, whose explicit form is checked by the fact that it properly removes all $1/\epsilon^n$ pole terms in the bare soft function at one-loop order, we stress again that the two-loop form of the anomalous dimension shown above should be considered as a conjecture. 

\subsection{Exact solution to the RG equation}
\label{sec:solu}

The RG equation (\ref{Sevol}) and the associated anomalous dimension $\gamma_S$ in (\ref{gammaS}) are closely related to the corresponding equations for the leading-twist LCDA $\phi_+^B(\omega)$ of the $B$ meson \cite{Grozin:1996pq,Lange:2003ff,Lee:2005gza}. This correspondence is discussed in more detail in Appendix~\ref{app:LCDA}. Structurally, the anomalous dimension $\gamma_S$ differs from the anomalous dimension for the LCDA by the argument of the logarithm ($w/\mu^2$ versus $\omega/\mu$, because $w$ has mass dimension 2 while $\omega$ has mass dimension 1) and a factor 2 in front of the non-local terms. In solving the evolution equation for the soft function we will use and extend some of the strategies developed in the literature on the $B$-meson LCDA. We will see, however, that the additional factor~2 in front of the non-local terms in the anomalous dimension give rise to major complications.

A closed analytic solution to the evolution equation (\ref{Sevol}) can be obtained based on the observation that, on dimensional grounds, any pure power $w^a$ provides an eigenfunction of the evolution kernel. To see this, note that the integrals \cite{Lange:2003ff}
\begin{equation}\label{calFdef}
   \F(a)\equiv \int_0^\infty\!dw'\,w\,\Gamma(w,w') \left( \frac{w'}{w} \right)^a
   = - \big[ H(a) + H(-a) \big] \,,
\end{equation}
where $H(a)=\psi(1+a)+\gamma_E$ is the harmonic-number function, and 
\begin{equation}
   \Ha(-a) \equiv \frac{1}{\beta_0} \int_0^\infty\!dw'\,\frac{w\,\theta(w'-w)}{w'(w'-w)}\,
    h\bigg(\frac{w}{w'}\bigg) \left( \frac{w'}{w} \right)^a 
\end{equation}
define two dimensionless functions of the exponent $a$, which are analytic in the complex $a$-plane with pole singularities at all positive and negative (for $\F$ only) integer values of $a$. Our choice of the function $\Ha(-a)$ is consistent with the definition 
\begin{equation}\label{Hdef}
\begin{aligned}
   \Ha(a) &= \frac{1}{\beta_0} \int_0^1\!\frac{dx}{1-x}\,h(x)\,x^a \\
   &= \left( \frac{3C_F}{\beta_0} -1 \right) \psi'(1+a) 
    + \frac{2C_F}{\beta_0} \left[ \frac{\psi'(1+a)}{a}
    - \left( \frac{1}{a^2} + 2\psi'(1+a) \right) H(a) \right]
\end{aligned}
\end{equation}
introduced in \cite{Liu:2020ydl}. It follows that the ansatz
\begin{equation}\label{ansatz}
\begin{aligned}
   & \left( \frac{w}{\mu_s^2} \right)^{\eta-a_\Gamma(\mu_s,\mu)} 
    \exp\Big[ 2 S(\mu_s,\mu) + a_{\gamma_s}(\mu_s,\mu) \Big] 
    \exp\Bigg[ 2 \int\limits_{\alpha_s(\mu_s)}^{\alpha_s(\mu)}\!d\alpha\,
    \frac{\Gamma_{\rm cusp}(\alpha)}{\beta(\alpha)}\,
    \F\big(\eta-a_\Gamma(\mu_s,\mu_\alpha)\big) \Bigg] \\[-2mm]
   &\quad \times \exp\Bigg[\,\int\limits_{\alpha_s(\mu_s)}^{\alpha_s(\mu)}\!\!
    \frac{d\alpha}{\beta(\alpha)}\,\bigg[ 2C_F \left( \frac{\alpha}{2\pi} \right)^2 
    \beta_0\,\Ha\big(a_\Gamma(\mu_s,\mu_\alpha)-\eta\big) + {\cal O}(\alpha^3) \bigg] \Bigg] 
\end{aligned}
\end{equation}
with $\alpha_s(\mu_\alpha)\equiv\alpha$ provides a solution to the RG equation (\ref{Sevol}) with the initial condition $\left(w/\mu_s^2\right)^\eta$ at some matching scale $\mu=\mu_s$, at which the soft function is free of large logarithms. Here $\beta(\alpha_s)=d\alpha_s/d\ln\mu$ is the QCD $\beta$-function, and we have introduced the RG functions (the first of which should not be confused with the soft function)
\begin{equation}\label{RGfuns}
\begin{aligned}
   S(\mu_s,\mu) 
   &= - \int\limits_{\alpha_s(\mu_s)}^{\alpha_s(\mu)}\!
    d\alpha\,\frac{\Gamma_{\rm cusp}(\alpha)}{\beta(\alpha)}
    \int\limits_{\alpha_s(\mu_s)}^\alpha
    \frac{d\alpha'}{\beta(\alpha')} \,, \\
   a_\Gamma(\mu_s,\mu) 
   &= - \int\limits_{\alpha_s(\mu_s)}^{\alpha_s(\mu)}\!
    d\alpha\,\frac{\Gamma_{\rm cusp}(\alpha)}{\beta(\alpha)} \,.
\end{aligned}
\end{equation}
They are the solutions to the equations 
\begin{equation}
\begin{aligned}
   \frac{d}{d\ln\mu}\,S(\mu_s,\mu) 
   &= - \Gamma_{\rm cusp}(\alpha_s)\,\ln\frac{\mu}{\mu_s} \,, \\
   \frac{d}{d\ln\mu}\,a_\Gamma(\mu_s,\mu) 
   & = - \Gamma_{\rm cusp}(\alpha_s) 
\end{aligned}
\end{equation}
with the boundary conditions $S(\mu_s,\mu_s)=0$ and $a_\Gamma(\mu_s,\mu_s)=0$. The function $a_{\gamma_s}(\mu_s,\mu)$ is defined analogously to $a_\Gamma(\mu_s,\mu)$. Note that both $S(\mu_s,\mu)$ and $a_\Gamma(\mu_s,\mu)$ take negative values if $\mu>\mu_s$, because the cusp anomalous dimension is a positive quantity. Explicit expressions for these objects obtained at NLO in RG-improved perturbation theory are given in Appendix~\ref{app:A}. 

Changing variables in the integral in the exponent in the first line of (\ref{ansatz}) from $\alpha$ to $\varrho=a_\Gamma(\mu_s,\mu_\alpha)$ allows us to perform this integral in closed form. Moreover, in the exponent of the term in the second line we can use that $\beta(\alpha_s)=-\beta_0\,\alpha_s^2/(2\pi)+\dots$ at leading order. We can thus rewrite (\ref{ansatz}) in the more compact form
\begin{equation}\label{ansatz2}
\begin{aligned}
   &U_S(w;\mu,\mu_s) \left( \frac{w}{\mu_s^2} \right)^\eta\, 
    \frac{\Gamma^2\big(1-\eta+a_\Gamma(\mu_s,\mu)\big)\,\Gamma^2(1+\eta)}%
         {\Gamma^2\big(1+\eta-a_\Gamma(\mu_s,\mu)\big)\,\Gamma^2(1-\eta)} \\
   &\times \exp\Bigg[ - C_F\!\!\int\limits_{\alpha_s(\mu_s)}^{\alpha_s(\mu)}\!
    \frac{d\alpha}{\pi}\,\Big[ \Ha\big(a_\Gamma(\mu_s,\mu_\alpha)-\eta\big) 
    + {\cal O}(\alpha) \Big] \Bigg] \,,
\end{aligned}
\end{equation}
where we have defined the evolution function
\begin{equation}\label{USdef}
   U_S(w;\mu,\mu_s) 
   = \left( \frac{w e^{-4\gamma_E}}{\mu_s^2} \right)^{-a_\Gamma(\mu_s,\mu)} 
    \exp\Big[ 2 S(\mu_s,\mu) + a_{\gamma_s}(\mu_s,\mu) \Big] \,, 
\end{equation}
which satisfies $U_S(w;\mu_s,\mu_s)=1$.

In order to apply this solution, we need to recast the initial condition for the soft function at the matching scale $\mu_s$ in such a way that it is written as a superposition of pure powers $\left(w/\mu_s^2\right)^\eta$. To this end, we express $S(w,\mu_s)$ in terms of its Laplace transform with respect to $\ln(w/m_b^2)$, which in general is defined as 
\begin{equation}\label{Stilde}
   \tilde S(\eta,\mu) = \int_0^\infty\!\frac{dw}{w}\,S(w,\mu)
    \left( \frac{w}{m_b^2} \right)^{-\eta} .
\end{equation}
In fixed-order perturbation theory the integral converges for $0<\eta<1$. We thus write
\begin{equation}\label{inverseLaplace}
   S(w,\mu_s) = \frac{1}{2\pi i} \int\limits_{c-i\infty}^{c+i\infty}\!d\eta\,\tilde S(\eta,\mu_s) 
    \left( \frac{w}{m_b^2} \right)^\eta ,
\end{equation}
with $0<c<1$. Since the right-hand side of (\ref{inverseLaplace}) exhibits a power-law dependence on $w$, we can use the solution (\ref{ansatz2}) to express the soft function at a different scale $\mu>\mu_s$ as
\begin{equation}\label{SoftsolL}
\begin{aligned}
   S(w,\mu) 
   &= U_S(w;\mu,\mu_s)\,\frac{1}{2\pi i} \int\limits_{c-i\infty}^{c+i\infty}\!d\eta\,
    \tilde S(\eta,\mu_s) \left( \frac{w}{m_b^2} \right)^\eta\, 
    \frac{\Gamma^2\big(1-\eta+a_\Gamma(\mu_s,\mu)\big)\,\Gamma^2(1+\eta)}%
         {\Gamma^2\big(1+\eta-a_\Gamma(\mu_s,\mu)\big)\,\Gamma^2(1-\eta)} \\
   &\quad \times \exp\Bigg[ - C_F\!\!\int\limits_{\alpha_s(\mu_s)}^{\alpha_s(\mu)}\!
    \frac{d\alpha}{\pi}\,\Ha\big(a_\Gamma(\mu_s,\mu_\alpha)-\eta\big) 
    + {\cal O}(\alpha_s^2) \Bigg] \,.
\end{aligned}
\end{equation}
In this expression the contour must be chosen such that $0<c<1+\min\big(0,a_\Gamma(\mu_s,\mu)\big)$, which in turn requires $a_\Gamma(\mu_s,\mu)>-1$.\footnote{This condition is always satisfied in practice. At leading order it is equivalent to the inequality $\alpha_s(\mu)>\exp\big(-\frac{\beta_0}{2C_F}\big)\,\alpha_s(\mu_s)\approx 0.056\,\alpha_s(\mu_s)$, where the numerical value refers to $n_f=5$ light-quark flavors.} 
Relation (\ref{SoftsolL}) provides an exact solution of the evolution equation (\ref{Sevol}) in the approximation where one ignores the yet unknown three-loop non-local terms in the soft anomalous dimension in (\ref{gammaS}), which would enter in the indicated higher-order corrections in the exponent of the last factor. However, this solution requires the Laplace transform $\tilde S(\eta,\mu_s)$ of the soft function at the matching scale $\mu_s$. 

Clearly, it would be more convenient to have a solution of the evolution equation that relates $S(w,\mu)$ in a more direct way with the initial condition $S(w,\mu_s)$. To obtain it, we use the fact that the exponent in the second line of (\ref{SoftsolL}) is of higher order in $\alpha_s$ and hence can be expanded out. Eliminating the Laplace transform $\tilde S(\eta,\mu_s)$ by means of (\ref{Stilde}) we then obtain
\begin{equation}\label{eq42}
\begin{aligned}
   S(w,\mu) 
   &= U_S(w;\mu,\mu_s)\,\int_0^\infty\!\frac{dw'}{w'}\,S(w',\mu_s)\,
    \frac{1}{2\pi i} \int\limits_{c-i\infty}^{c+i\infty}\!d\eta \left( \frac{w}{w'} \right)^\eta 
    \frac{\Gamma^2\big(1-\eta+a_\Gamma(\mu_s,\mu)\big)\,\Gamma^2(1+\eta)}%
         {\Gamma^2\big(1+\eta-a_\Gamma(\mu_s,\mu)\big)\,\Gamma^2(1-\eta)} \\
   &\quad\times \Bigg[ 1 - \frac{C_F}{\beta_0\pi}
    \int\limits_{\alpha_s(\mu_s)}^{\alpha_s(\mu)}\!d\alpha \int_0^1\!\frac{dx}{1-x}\,h(x)\,
    x^{a_\Gamma(\mu_s,\mu_\alpha)-\eta} + {\cal O}(\alpha_s^2) \Bigg] \,,
\end{aligned}
\end{equation}
where we have used the definition of the function $\Ha(a)$ in (\ref{Hdef}). The integrand of the integral over $\eta$ has single and double poles located at $\eta=n+a_\Gamma(\mu_s,\mu)$ and $\eta=-n$ for $n\in\mathbb{N}$. As long as $a_\Gamma(\mu_s,\mu)>-1$, these two series of poles lie on different sides of the integration contour if we choose $0<c<1+a_\Gamma(\mu_s,\mu)$. The integral can then be expressed in terms of Meijer $G$-functions (see e.g.\ \cite{Gfun2,Gfun3} for a detailed discussion). This yields 
\begin{equation}\label{eq42b}
\begin{aligned}
   S(w,\mu) 
   &= U_S(w;\mu,\mu_s)\,\int_0^\infty\!\frac{dw'}{w'}\,S(w',\mu_s)\,\Bigg[
    \MeijerG[\bigg]{2}{2}{4}{4}{-a,\, -a,\, 1-a,\, 1-a}{1,\, 1,\, 0,\, 0}{\frac{w'}{w}} \\
   &\quad\mbox{}- \frac{C_F}{\beta_0\pi}\,\int_0^1\!\frac{dx}{1-x}\,h(x) 
    \int\limits_{\alpha_s(\mu_s)}^{\alpha_s(\mu)}\!d\alpha\,x^{a_\Gamma(\mu_s,\mu_\alpha)}\, 
    \MeijerG[\bigg]{2}{2}{4}{4}{-a,\, -a,\, 1-a,\, 1-a}{1,\, 1,\, 0,\, 0}{\frac{x w'}{w}} 
    + {\cal O}(\alpha_s^2) \Bigg] \,,
\end{aligned}
\end{equation}
where $a=a_\Gamma(\mu_s,\mu)$. This form of the solution is valid as long as one considers an upward scale evolution ($\mu>\mu_s$), for which $a<0$. The Meijer $G$-function appearing above becomes singular when the last argument approaches 1, 
\begin{equation}\label{singular}
   \lim_{z\to 1}\,\MeijerG[\bigg]{2}{2}{4}{4}{-a,\, -a,\, 1-a,\, 1-a}{1,\, 1,\, 0,\, 0}{z} 
   = - \frac{\sin2\pi a}{\pi}\,\frac{\Gamma(1+4a)}{\left| 1-z \right|^{1+4a}} 
    + {\cal O}(1) \,.
\end{equation}
This is an integrable singularity only as long as $a<0$. It is possible to relate the Meijer $G$-function to an expression involving a hypergeometric function and its derivatives by performing the contour integral in (\ref{eq42}) using the theorem of residues. This is discussed in more detail in Appendix~\ref{app:B}.

There are two further simplifications that can be made, because the second term inside the brackets in relation (\ref{eq42b}) is of ${\cal O}(\alpha_s)$. For this term it is thus sufficient to use the leading-order expressions for the soft function and for the RG function
\begin{equation}\label{eq41}
   a_\Gamma(\mu_s,\mu_\alpha) = \frac{2C_F}{\beta_0} \ln\frac{\alpha}{\alpha_s(\mu_s)}
    + {\cal O}(\alpha_s) \,,
\end{equation}
where $\beta_0=\frac{23}{3}$ is the one-loop coefficient of the $\beta$-function evaluated with $n_f=5$ light quark flavors (see Appendix~\ref{app:A}). We then find the final solution
\begin{equation}\label{boundlessbeauty}
\begin{aligned}
   S(w,\mu) 
   &= U_S(w;\mu,\mu_s)\,\Bigg[ \int_0^\infty\!\frac{dx}{x}\,S(w/x,\mu_s)\,
    \MeijerG[\bigg]{2}{2}{4}{4}{-a,\, -a,\, 1-a,\, 1-a}{1,\, 1,\, 0,\, 0}{\frac{1}{x}} \\
   &\quad\mbox{}- m_b\,\frac{C_F\alpha_s(\mu_s)}{\pi} \int_0^1\!\frac{dx}{1-x}\,\frac{h(x)}{\beta_0}\,
    \frac{r^{1+\frac{2C_F}{\beta_0} \ln x}-1}{1+\frac{2C_F}{\beta_0} \ln x}\,
    \MeijerG[\bigg]{2}{2}{4}{4}{-a,\, -a,\, 1-a,\, 1-a}{0,\, 1,\, 0,\, 0}{\frac{x m_b^2}{w}} 
    + {\cal O}(\alpha_s^2) \Bigg] ,   
\end{aligned}
\end{equation}
with $r=\alpha_s(\mu)/\alpha_s(\mu_s)$ and $a=a_\Gamma(\mu_s,\mu)$. The Meijer $G$-function enjoys the properties
\begin{equation}
   \MeijerG[\bigg]{2}{2}{4}{4}{-a,\, -a,\, 1-a,\, 1-a}{1,\, 1,\, 0,\, 0}{\frac{1}{z}} 
   = z^a\,\MeijerG[\Big]{2}{2}{4}{4}{-a,\, -a,\, 1-a,\, 1-a}{1,\, 1,\, 0,\, 0}{z}
   = \MeijerG[\Big]{2}{2}{4}{4}{0,\, 0,\, 1,\, 1}{1+a,\, 1+a,\, a,\, a}{z} \,,
\end{equation}
which can be used to restrict the last argument to values in the interval $[0,1]$. This is particularly useful for a numerical evaluation. 

Relation (\ref{boundlessbeauty}) provides the desired result for the exact evolution of the soft function from the matching scale $\mu_s$ to a different scale $\mu$ at NLO in RG-improved perturbation theory. In evaluating this result consistently one should use the one-loop expression (\ref{softrenorm}) for the soft function $S(w,\mu_s)$ at the matching scale, the two-loop expression for the anomalous dimension $\gamma_s$ and the three-loop expression for the cusp anomalous dimension (see Appendix~\ref{app:A}). In the literature on double logarithmic resummations this approximation is often referred to as the next-to-next-to-leading logarithmic (NNLL) approximation. It correctly resums logarithms of the form $\alpha_s^n L^k$ with $2n-3\le k\le 2n$ \cite{Becher:2007ty}.

The above solution simplifies considerably if one works at leading order in RG-improved perturbation theory, corresponding to the NLL approximation, which resums the logarithms with $2n-1\le k\le 2n$. One then uses the two-loop approximation for the cusp anomalous dimension, the one-loop approximation for $\gamma_s$ and the tree-level matching condition for the soft function at the scale $\mu_s$. In this approximation the integral over the Meijer $G$-function in (\ref{boundlessbeauty}) can be performed analytically, and we find
\begin{equation}\label{RGEsolLO}
   S_{\rm LO}(w,\mu) = m_b\,U_S(w;\mu,\mu_s)\, 
    \MeijerG[\bigg]{2}{2}{4}{4}{-a,\, -a,\, 1-a,\, 1-a}{0,\, 1,\, 0,\, 0}{\frac{m_b^2}{w}} \,;
    \quad a=a_\Gamma(\mu_s,\mu) \,.
\end{equation}

\begin{figure}
\begin{center}
\includegraphics[width=0.49\textwidth]{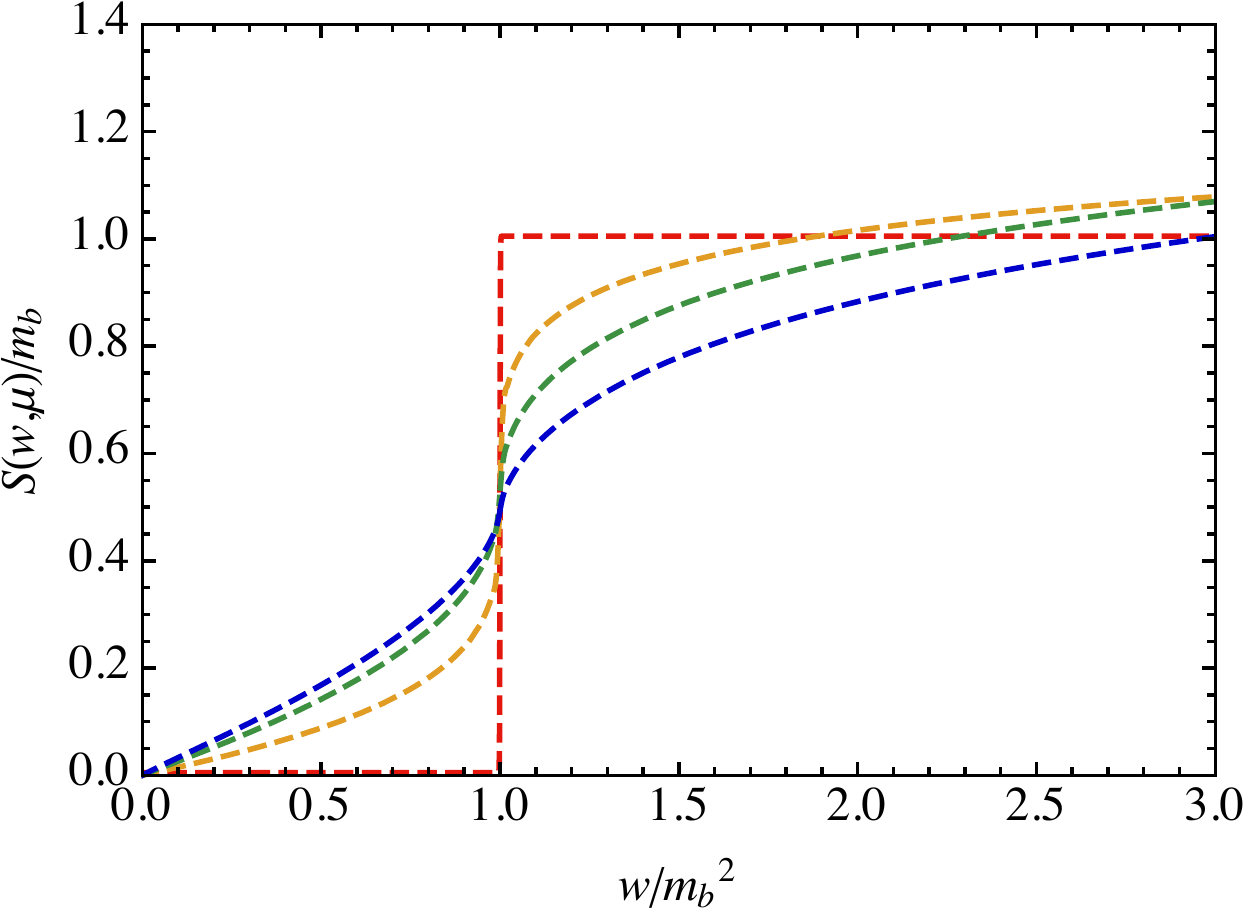} 
\includegraphics[width=0.49\textwidth]{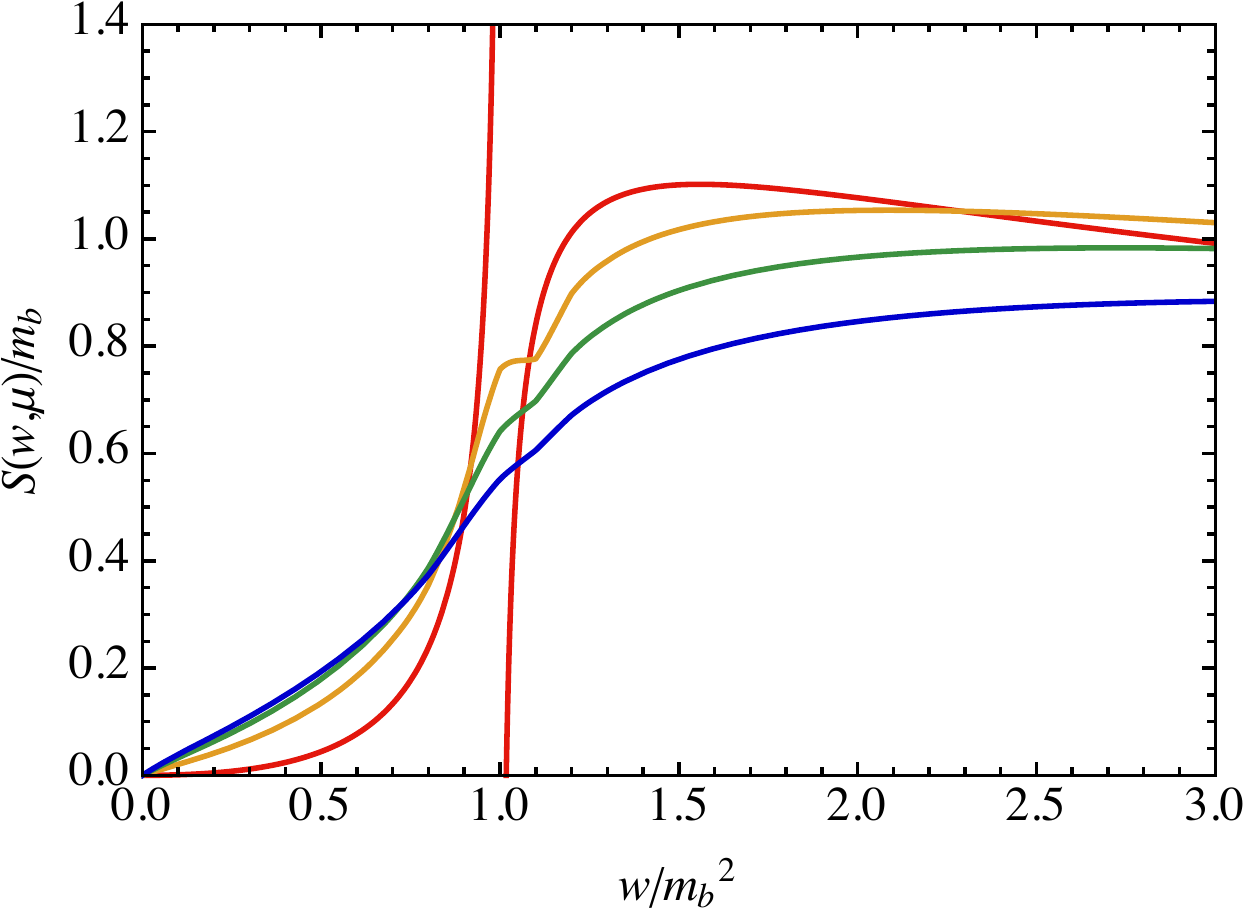}
\vspace{2mm}
\caption{\label{fig:Sevol} 
Renormalized soft function $S(w,\mu)/m_b$ for $\mu=m_b$ (red), 10\,GeV (orange), 20\,GeV (green) and 40\,GeV (blue), at leading order (left) and NLO (right) in RG-improved perturbation theory. The soft matching scale is set to $\mu_s=m_b$ in all cases.} 
\end{center}
\end{figure}

In Figure~\ref{fig:Sevol} we show the renormalized soft function $S(w,\mu)$ at different values of the renormalization scale $\mu$ and fixed matching scale $\mu_s=m_b=4.8$\,GeV. These solutions are derived by evolving the fixed-order expression at the matching scale $\mu_s=m_b$ obtained from (\ref{softrenorm}), and shown in Figure~\ref{fig:S1loop}, to the scale $\mu$ using the leading-order and NLO solutions to the RG equation presented in (\ref{RGEsolLO}) and (\ref{boundlessbeauty}), respectively. Note that the discontinuous behavior of the soft function at $w=m_b^2$ present in fixed-order perturbation theory (see Figure~\ref{fig:S1loop}) has been smoothed out after RG evolution to a higher scale. However, for relatively low values of~$\mu$ a prominent feature near $w=m_b^2$ remains.

\section{Asymptotic behavior and dynamical scale setting}
\label{sec:6}

In the solutions discussed in the previous section it is important that the matching scale $\mu_s$ is chosen such that the initial condition $S(w,\mu_s)$ for the soft function is free of large logarithms. At one-loop order and beyond, the explicit expression for this initial condition involves the logarithms $L_w=\ln(w/\mu^2)$ and $L_m=\ln(m_b^2/\mu^2)$, see (\ref{eqs18}). The RG-improved solutions for the soft function are thus well defined as long as $\mu_s\sim m_b\sim\sqrt{w}$ are all of the same order. Indeed, in Figure~\ref{fig:Sevol} we have set $\mu_s=m_b$ and varied $w$ up to a maximum value such that $\sqrt{w}<\sqrt{3}\,m_b$. Clearly, it would be desirable to be able to treat the two soft scales $m_b$ and $w$ as independent and extend the solution for the soft function up to values $w\gg m_b^2$. We now show how this can be accomplished.

We have seen in (\ref{ZS1loop}) and (\ref{gammaS}) that the renormalization factor and the anomalous dimension for the soft function do not contain any reference to the $b$-quark mass. Apart from the prefactor of $m_b$ in the normalization of the soft function, nothing prevents us from studying this function and its RG evolution in the limit $m_b\to 0$ or, more properly, $w/m_b^2\to\infty$. Denoting the function in this limit by $S_\infty(w,\mu)$, we find from (\ref{softrenorm}) (with $w>0$) 
\begin{equation}
\begin{aligned}
   S_\infty(w,\mu) 
   &= m_b \left[ 1 + \frac{C_F\alpha_s}{4\pi} 
    \left( - L_w^2 - 6 L_w + 3 L_m + 8 - \frac{\pi^2}{2} \right) + {\cal O}(\alpha_s^2) \right] \\
   &= m_b(\mu) \left[ 1 + \frac{C_F\alpha_s}{4\pi} 
    \left( - L_w^2 - 6 L_w + 12 - \frac{\pi^2}{2} \right) + {\cal O}(\alpha_s^2) \right] .
\end{aligned}
\end{equation}
One subtle point is that in this limit one should use the running $b$-quark mass $m_b(\mu)$ in the prefactor, as shown in the second line. The reason is that the pole mass contains a logarithm $L_m$ in its definition, 
\begin{equation}
   m_b = m_b(\mu) \left[ 1 + \frac{C_F\alpha_s}{4\pi} 
    \left( - 3 L_m + 4 \right) + {\cal O}(\alpha_s^2) \right] ,
\end{equation}
and hence the limit $m_b\to 0$ would be singular. Once this is done, the higher-order corrections to the soft function in the limit of large $w\gg m_b^2$ only contain powers of the logarithm $L_w$ and constants. This fact greatly simplifies the solution of its RG evolution equation. In a first step, we define a new function $\Sa_\infty(L_w,\mu)$ via 
\begin{equation}
   S_\infty(w,\mu)\equiv m_b(\mu)\,\Sa_\infty(L_w,\mu) \,.
\end{equation}
From (\ref{ansatz2}) it follows that the general solution to the evolution equation for $S_\infty(w,\mu)$ can be written in the form \cite{Liu:2020ydl} 
\begin{equation}\label{Sinftysol}
\begin{aligned}
   S_\infty(w,\mu) 
   &= m_b(\mu_s)\,U_S(w;\mu,\mu_s)\,\,\Sa_\infty(\partial_\eta,\mu_s)
    \left( \frac{w}{\mu_s^2} \right)^\eta\, 
    \frac{\Gamma^2\big(1-\eta+a_\Gamma(\mu_s,\mu)\big)\,\Gamma^2(1+\eta)}%
         {\Gamma^2\big(1+\eta-a_\Gamma(\mu_s,\mu)\big)\,\Gamma^2(1-\eta)} \\
   &\quad\times \Bigg[ 1 - C_F\!\!\int\limits_{\alpha_s(\mu_s)}^{\alpha_s(\mu)}\!
    \frac{d\alpha}{\pi}\,\Ha\big(a_\Gamma(\mu_s,\mu_\alpha)-\eta\big) 
    + {\cal O}(\alpha_s^2) \Bigg]\,\Bigg|_{\eta=0} \,.
\end{aligned}
\end{equation}
At NLO in RG-improved perturbation theory one can set $\eta=0$ in the argument of the function $\Ha$, because the tree-level term in $\Sa_\infty(L_w,\mu_s)$ is a constant. In this expression, which is indeed much simpler than the exact solution (\ref{boundlessbeauty}), one needs the explicit form of the function $\Ha(a)$ given in (\ref{Hdef}). 

The NLO solutions (\ref{boundlessbeauty}) and (\ref{Sinftysol}) are formally independent of the matching scale $\mu_s$. In the limit where $w\gg m_b^2$ it is inconsistent to evaluate them with a fixed scale choice $\mu_s={\cal O}(m_b)$. Rather, one should make a {\em dynamical scale choice}, such that $\mu_s^2=r_s w$ with $r_s={\cal O}(1)$. In the asymptotic solution (\ref{Sinftysol}) the dependence on $w$ then enters via the running coupling $\alpha_s(\mu_s)$ only. Note that for a fixed value of the renormalization scale $\mu$ there is always a range of $w$ values for which $\mu_s>\mu$. It is important in this context that the asymptotic solution is not restricted to the case where $\mu>\mu_s$, unlike the solution in (\ref{boundlessbeauty}). 

\begin{figure}
\begin{center}
\includegraphics[width=0.48\textwidth]{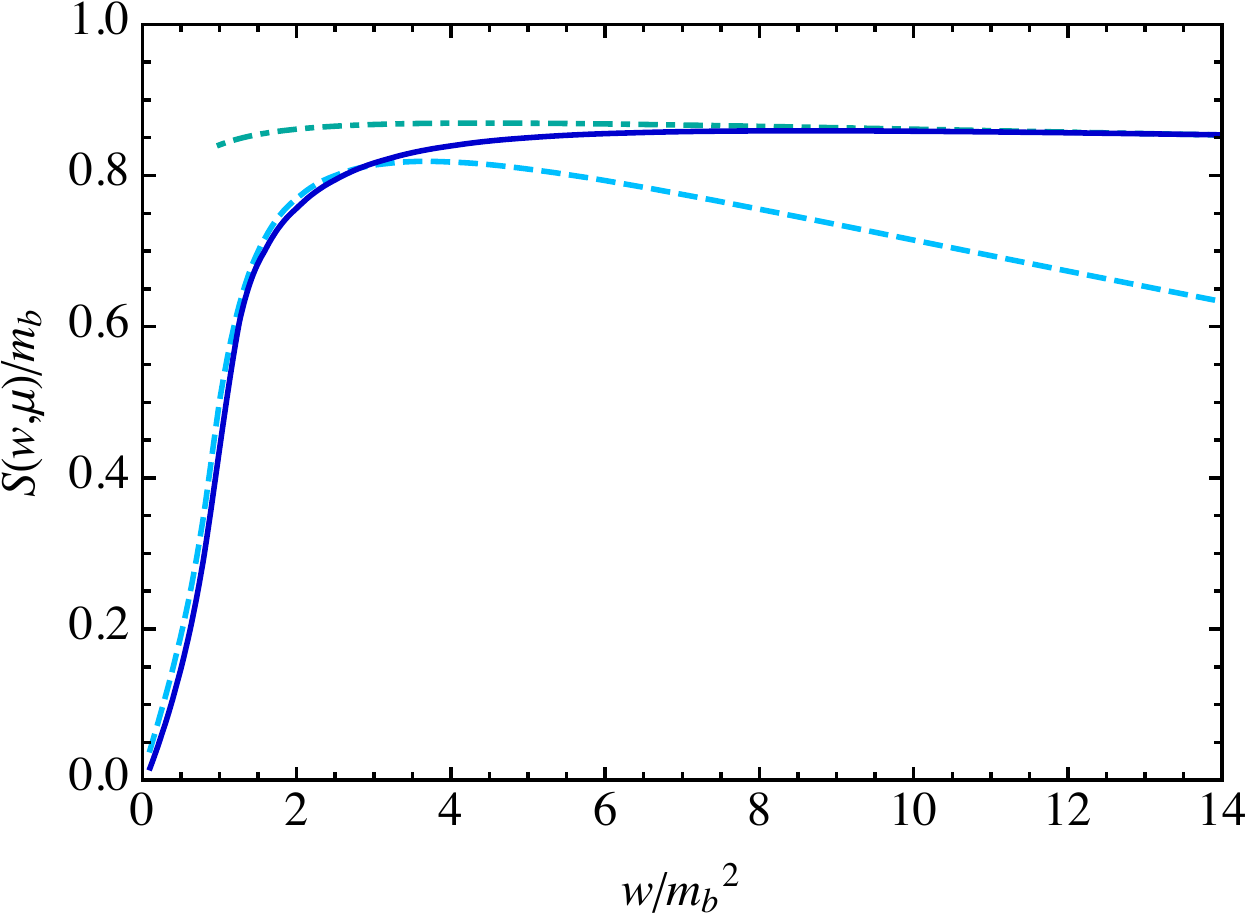}\quad
\includegraphics[width=0.48\textwidth]{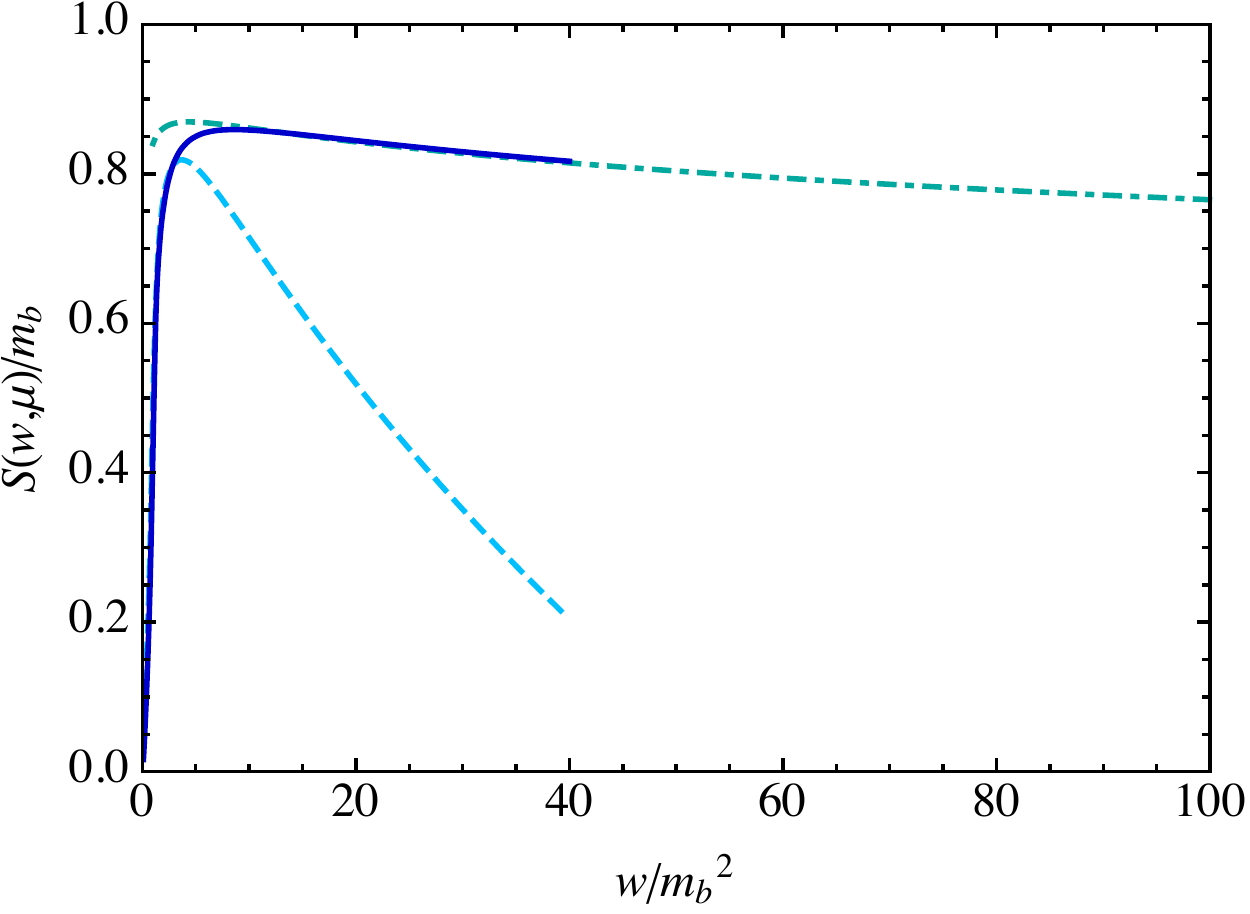}
\vspace{2mm}
\caption{\label{fig:Sasy} 
Different solutions for the soft function $S(w,\mu)$ obtained at NLO in RG-improved perturbation theory. The renormalization scale is fixed at $\mu=60$\,GeV. The solid line shows the result (\ref{boundlessbeauty}) evaluated with the dynamical scale choice $\mu_s=\mbox{max}(m_b,\sqrt{w})$. The dashed line corresponds to the same solution with fixed $\mu_s=m_b$. The dash-dotted curve shows the asymptotic solution (\ref{Sinftysol}) with $\mu_s=\sqrt{w}$. The right panel illustrates the behavior at larger values of $w$.}
\end{center}
\end{figure}

Figure~\ref{fig:Sasy} shows different results for the soft function $S(w,\mu)$ obtained at NLO in RG-improved perturbation theory. The renormalization scale is fixed to $\mu=60$\,GeV. We still normalize our results to the $b$-quark pole mass, such that the plots can be compared with those in Figure~\ref{fig:Sevol}. We evaluate the ratio $m_b(\mu)/m_b$ at NLO in RG-improved perturbation theory (see Appendix~\ref{app:A} for details). The dashed curve shows our previous solution computed from (\ref{boundlessbeauty}) with the fixed scale choice $\mu_s=m_b$. The dash-dotted curve shows the asymptotic solution for $w\gg m_b$ obtained from (\ref{Sinftysol}) by setting $\mu_s=\sqrt{w}$. The solid curve shows the solution derived from (\ref{boundlessbeauty}) using the dynamical scale choice $\mu_s=\mbox{max}(m_b,\sqrt{w})$. To obtain it, we rewrite (\ref{softrenorm}) by pulling out the running mass $m_b(\mu)$ rather than the pole mass. This has the effect of replacing the term $(3L_m+8)$ in the expression for the function $S_a(w,\mu)$ in (\ref{eqs18}) by 12. The fact that there is no such compensating effect for the function $S_b(w,\mu)$, which starts at ${\cal O}(\alpha_s)$, explains the small difference between the solid and dashed lines in the region of small $w$. We then use this modified form for the matching condition $S(w,\mu_s)$. The solid curve provides the optimal RG-improved solution for the soft function valid for small and large values of $w$. Note, however, that in practice we cannot evaluate this solution for arbitrarily large $w$ values, because this would violate the condition $\mu_s<\mu$, which is a prerequisite for the analytic solution shown in (\ref{boundlessbeauty}) to hold. We observe that the solution with the fixed choice $\mu_s=m_b$ of the matching scale breaks down for $w>3\hspace{0.3mm}m_b^2$, because the ``large logarithms'' $\ln(w/\mu_s^2)$ arising in this region are not resummed in this case. The asymptotic solution (\ref{Sinftysol}) provides an accurate description for $w>10\hspace{0.3mm}m_b^2$, and it can be evaluated for arbitrarily large values of $w$. The right panel in the figure shows the solution up to $w=(10\hspace{0.3mm}m_b)^2$.

\section{Soft function in Laplace space}
\label{sec:Laplace}

A much simpler solution to the RG evolution of the soft function is found in Laplace space. We define the Laplace transform $\tilde S(\eta,\mu)$ of the soft function at the scale $\mu$ as shown in (\ref{Stilde}). It then follows from (\ref{SoftsolL}) that
\begin{equation}\label{verynice}
\begin{aligned}
   \tilde S(\eta,\mu) 
   &= U_S(m_b^2;\mu,\mu_s)\, 
    \frac{\Gamma^2\big(1+\eta+a_\Gamma(\mu_s,\mu)\big)\,\Gamma^2(1-\eta)}%
         {\Gamma^2\big(1-\eta-a_\Gamma(\mu_s,\mu)\big)\,\Gamma^2(1+\eta)} \\
   &\quad \times \exp\Bigg[ - C_F\!\!\int\limits_{\alpha_s(\mu_s)}^{\alpha_s(\mu)}\!
    \frac{d\alpha}{\pi}\,\Ha\big(-\eta+a_\Gamma(\mu,\mu_\alpha)\big) 
    + {\cal O}(\alpha_s^2) \Bigg]\,\tilde S\big(\eta+a_\Gamma(\mu_s,\mu),\mu_s\big) \,.
\end{aligned}
\end{equation}
Contrary to (\ref{eq42b}) no integral over the soft function is required. Instead, under a scale transformation the argument of the Laplace transform is shifted by an amount $a_\Gamma(\mu_s,\mu)$. Recall that the Laplace transform $\tilde S(\eta,\mu_s)$ of the soft function at the matching scale $\mu_s$ exists for $0<\eta<1$, as is evident from the explicit formula shown in (\ref{1loopLaplace}) below. It then follows from the above result that the Laplace transform at a higher scale $\mu>\mu_s$ exists for $-a_\Gamma(\mu_s,\mu)=|a_\Gamma(\mu_s,\mu)|<\eta<1$.

It is not difficult to check that (\ref{verynice}) is the solution of the partial differential equation 
\begin{equation}\label{RGELaplace}
\begin{aligned}
   & \left( \frac{d}{d\ln\mu} 
    + \Gamma_{\rm cusp}(\alpha_s)\,\frac{\partial}{\partial\eta} \right) \tilde S(\eta,\mu) \\
   &= \bigg[ \Gamma_{\rm cusp}(\alpha_s) \left( \ln\frac{m_b^2}{\mu^2} + 2\F(-\eta) \right)
    - \gamma_s(\alpha_s) + 2 C_F\beta_0 \left( \frac{\alpha_s}{2\pi} \right)^2 \Ha(-\eta)
    + {\cal O}(\alpha_s^3) \bigg]\,\tilde S(\eta,\mu) \,,
\end{aligned}
\end{equation}
where the functions $\F(a)$ and $\Ha(a)$ have been defined in (\ref{calFdef}) and (\ref{Hdef}), respectively. This generalized RG evolution equation for $\tilde S(\eta,\mu)$ can be derived from relation (\ref{inverseLaplace}), with $\mu_s$ replaced by $\mu$, and the RG equation for the soft function following from (\ref{Sevol}) and (\ref{gammaS}). 

The Laplace transform $\tilde S(\eta,\mu_s)$ at the matching scale $\mu_s$ can be derived from the NLO expression for the soft function given in (\ref{softrenorm}). At one-loop order we obtain \begin{equation}\label{1loopLaplace}
\begin{aligned}
   \tilde S(\eta,\mu_s) 
   &= \frac{m_b}{\eta}\,\Bigg\{ 1 + \frac{C_F\alpha_s(\mu_s)}{4\pi}\,\bigg[
    - \frac{2}{\eta^2} - \frac{2}{\eta} \left( L_m+3 \right) 
    - L_m^2 - 3 L_m + 8 - \frac{3\pi^2}{2} \\
   &\hspace{4.35cm}\mbox{}+ 4 L_m \big[ H(\eta) + H(-\eta) \big] 
    + \frac{4(1+\eta)\,H(\eta)}{\eta} \\
   &\hspace{4.35cm}\mbox{}- 4 \big[ H^2(\eta) + H^2(-\eta) \big] 
    - 2\psi'(1+\eta) + 4\psi'(1-\eta) \bigg] \Bigg\} \,.
\end{aligned}
\end{equation}
As before $H(a)$ is the harmonic-number function and $L_m=\ln(m_b^2/\mu_s^2)$. Given this result, it is straightforward to work out the NLO solution for $\tilde S(\eta,\mu)$ at a different scale from (\ref{verynice}).

The solution for the Laplace transform of the soft function derived here is more than just an academic result. In fact, it is well known that many factorization theorems involving complicated convolutions of jet and soft functions take on a particularly simple form in Laplace space (see e.g.\ \cite{Neubert:2005nt,Becher:2006nr}). The reason is that after RG improvement the jet functions $J(p^2)$ can often be written as a derivative operator acting on a factor $\left(p^2/\mu_j^2\right)^a$ with some exponent $a$. Thus the convolution of one or more such jet functions with a soft function evaluates naturally to the Laplace transform of the soft function.

\section{Soft function in the diagonal space}
\label{sec:dual}

It has been shown in \cite{Bell:2013tfa,Braun:2014owa} that one can bring RG evolution equations of the type (\ref{Sevol}) -- in the approximation where the non-local two-loop contributions in the anomalous dimension $\gamma_S$ shown in the second line of (\ref{gammaS}) are neglected -- into a simpler form using a suitably chosen integral transformation. The so-called ``dual'' soft function $s(w,\mu)$ obtained after this transformation obeys an evolution equation that is {\em local\/} in the variable $w$ and hence much easier to solve. In the context of our problem, this equation would take the form
\begin{equation}\label{localRGE}
   \frac{d}{d\ln\mu}\,s(w,\mu) 
   = \left[ \Gamma_{\rm cusp}(\alpha_s)\,\ln\frac{w e^{-4\gamma_E}}{\mu^2}
    - \gamma_s(\alpha_s) \right] s(w,\mu) \,.
\end{equation}
We discuss the construction of the dual soft function in the above-stated approximation in Appendix~\ref{app:D}. It requires a straightforward extension of the technology developed in \cite{Bell:2013tfa}.\footnote{The use of the word ``dual'' in this context is misleading. The Laplace variable $\eta$ is dual to the momentum-space variable $\ln(w/m_b^2)$ in the sense that small $\eta$ corresponds to large $w$ and vice versa. On the contrary, the variables $w$ in momentum space and in the dual space have similar physical meaning (as the square of a soft momentum scale).}

When the non-local two-loop contributions neglected above are put back in place, we find that the dual soft function obeys the RG equation
\begin{equation}\label{RGEdual}
   \frac{d}{d\ln\mu}\,s_{\rm dual}(w,\mu) 
   = - \int_0^\infty\!dw'\,\gamma_S^{\rm dual}(w,w';\mu)\,s_{\rm dual}(w',\mu)
\end{equation}
with
\begin{equation}\label{gammaSdual}
\begin{aligned}
   \gamma_S^{\rm dual}(w,w';\mu) 
   &= - \left[ \Gamma_{\rm cusp}(\alpha_s)\,
    \ln\frac{w e^{-4\gamma_E}}{\mu^2} - \gamma_s(\alpha_s) \right] \delta(w-w') \\
    &\quad\mbox{}- 2 C_F \left( \frac{\alpha_s}{2\pi} \right)^2 
    \frac{w\,\theta(w'-w)}{w'(w'-w)}\,h\bigg(\frac{w}{w'}\bigg) + {\cal O}(\alpha_s^3) \,.
\end{aligned}
\end{equation}
This follows from the analogy with the case of the $B$-meson LCDA discussed in Appendix~\ref{app:LCDA}, for which the two-loop anomalous dimension in the dual space has been derived in \cite{Braun:2019wyx}. Equation (\ref{gammaSdual}) is simpler than (\ref{gammaS}), because the term proportional to the plus distribution $\Gamma(w,w')$ has been eliminated. However, starting at two-loop order the dual anomalous dimension contains non-local terms with $w\ne w'$, thus upsetting the original motivation for the construction of the dual space. The complexity of the original evolution equation is therefore reduced only marginally. In this sense the approach of \cite{Bell:2013tfa,Braun:2014owa} leads to a hybrid representation, in which the one-loop anomalous-dimension kernel is diagonalized but higher-order corrections to it are not. We will now generalize the approach of \cite{Bell:2013tfa,Braun:2014owa} and construct what we refer to as the ``diagonal space'', in which the RG equation for the soft function retains the simple local form (\ref{localRGE}) to all orders of perturbation theory. 

\subsection{Construction of the diagonal space}

The key observation is based on the structure of the solution to the RG equation for the Laplace transform of the soft function given in (\ref{verynice}). Shifting the Laplace variable $\eta$ by $-a_\Gamma(\mu_s,\mu)$ we can rearrange this solution in the form
\begin{equation}\label{Laplacesplit}
\begin{aligned}
   & \frac{\Gamma^2\big(1+\eta-a_\Gamma(\mu_s,\mu)\big)}%
          {\Gamma^2\big(1-\eta+a_\Gamma(\mu_s,\mu)\big)}\,
    \exp\Bigg[ - C_F\!\!\int\limits_{\alpha_s(\mu)}^{\alpha_s(\rho)}\!\frac{d\alpha}{\pi}\,
    \Ha\big(-\eta+a_\Gamma(\mu_s,\mu)+a_\Gamma(\mu,\mu_\alpha)\big) 
    + {\cal O}(\alpha_s^2) \Bigg] \\
   &\times \tilde S\big(\eta-a_\Gamma(\mu_s,\mu),\mu\big) \\
   &= U_S(m_b^2;\mu,\mu_s)\,\frac{\Gamma^2(1+\eta)}{\Gamma^2(1-\eta)} 
    \exp\Bigg[ - C_F\!\!\int\limits_{\alpha_s(\mu_s)}^{\alpha_s(\rho)}\!
    \frac{d\alpha}{\pi}\,\Ha\big(-\eta+a_\Gamma(\mu_s,\mu_\alpha)\big) + {\cal O}(\alpha_s^2) \Bigg]\,
    \tilde S(\eta,\mu_s) \,,
\end{aligned}
\end{equation}
where we have used that $a_\Gamma(\mu_s,\mu)+a_\Gamma(\mu,\mu_\alpha)=a_\Gamma(\mu_s,\mu_\alpha)$. The auxiliary scale parameter $\rho$ is arbitrary and only serves to split up the integral into two terms. It follows that the function
\begin{equation}\label{gtdef}
   g(w,\eta,\mu,\rho)\equiv \frac{\Gamma^2(1+\eta)}{\Gamma^2(1-\eta)}\,
    \exp\Bigg[ - C_F\!\!\int\limits_{\alpha_s(\mu)}^{\alpha_s(\rho)}\!\frac{d\alpha}{\pi}\,
    \Ha\big(-\eta+a_\Gamma(\mu,\mu_\alpha)\big) + {\cal O}(\alpha_s^2) \Bigg]\,
    \tilde S(\eta,\mu) \left( \frac{w}{m_b^2} \right)^\eta ,
\end{equation}
has a particularly simple behavior under RG evolution, namely
\begin{equation}
   g\big(w,\eta-a_\Gamma(\mu_s,\mu),\mu,\rho\big) 
   = U_S(w;\mu,\mu_s)\,g(w,\eta,\mu_s,\rho) \,.
\end{equation}
Under scale evolution from $\mu_s$ to $\mu$ the second argument of $g$ gets shifted and the entire function is rescaled by the $\eta$-independent factor $U_S$. If we integrate both sides of this equation over a suitably chosen contour in the complex $\eta$-plane, which runs parallel to the imaginary axis with $\text{Re}(\eta)=c$ in the interval $0<c<1+\min\big(0,a_\Gamma(\mu_s,\rho)\big)$, then the shift becomes unobservable. Indeed, defining the soft function in the diagonal space via the inverse Laplace transform 
\begin{equation}\label{sdual}
   s(w,\mu,\rho) = \frac{1}{2\pi i} \int\limits_{c-i\infty}^{c+i\infty}\!d\eta\,g(w,\eta,\mu,\rho) \,,
\end{equation}
where $-a_\Gamma(\mu_s,\mu)<c<1+\min\big(0,a_\Gamma(\mu,\rho)\big)$, we obtain the simple result
\begin{equation}\label{sLOsol}
   s(w,\mu,\rho) = U_S(w;\mu,\mu_s)\,s(w,\mu_s,\rho) \,.
\end{equation}
This function obeys the local RG equation (\ref{localRGE}) to all orders in perturbation theory.\footnote{It would be possible to avoid the occurrences of exponentials of $\gamma_E$ in relations such as (\ref{localRGE}) by inserting the factor $e^{4\gamma_E\eta}$ under the integral on the right-hand side of (\ref{sdual}), thereby rescaling the $w$ variable in the diagonal space by $e^{4\gamma_E}$. This would be analogous to the transition from the MS to the $\overline{\rm MS}$ subtraction scheme.} 
In our discussion we have only included the non-local two-loop terms in the anomalous dimension~(\ref{gammaS}), but the method described here can readily be extended to higher orders. When terms beyond the two-loop order are taken into account, one just needs to add the corresponding higher-order corrections in the exponential in (\ref{gtdef}).

The soft function in the diagonal space depends on the auxiliary factorization scale $\rho$ in addition to the renormalization scale $\mu$. While its evolution in $\mu$ is local, as shown in (\ref{localRGE}), the evolution in $\rho$ is more complicated. From (\ref{gtdef})
it follows that
\begin{equation}\label{dlogrhoS}
   \frac{d}{d\ln\rho}\,s(w,\mu,\rho)
   = \bigg[ 2 C_F \left( \frac{\alpha_s(\rho)}{2\pi} \right)^2 \int_0^1\!\frac{dx}{1-x}\,
    h(x)\,x^{a_\Gamma(\mu,\rho)} + {\cal O}(\alpha_s^3) \bigg]\,s(w/x,\mu,\rho) \,.
\end{equation}
In all physical quantities the dependence on the scale $\rho$ drops out. Note that logarithms of the scale ratio $\rho/\mu$ are already resummed in this expression, so there is no need to choose these two scales to be of the same order.

\subsection{Transformation between momentum space and diagonal space} 
\label{sec:trafo}

Without loss of generality we write the linear relations connecting the original soft function $S(w,\mu)$ with the soft function $s(w,\mu,\rho)$ in the diagonal space in the general form
\begin{equation}\label{Strafo}
\begin{aligned}
   s(w,\mu,\rho) 
   &= \int_0^\infty\!\frac{dx}{\sqrt{x}}\,F_S(x,\mu,\rho)\,S(x w,\mu) \,, \\
   S(w,\mu) 
   &= \int_0^\infty\!\frac{dx}{\sqrt{x}}\,F_S^{\rm inv}(x,\mu,\rho)\,s(w/x,\mu,\rho) \,,
\end{aligned}
\end{equation}
where the two transfer functions $F_S$ and $F_S^{\rm inv}$ explicitly depend on the renormalization scale $\mu$ and on the auxiliary scale $\rho$. Combining (\ref{sdual}), (\ref{gtdef}) and (\ref{Stilde}) we find that
\begin{equation}\label{FSdef}
   \sqrt{x}\,F_S(x,\mu,\rho)
   = \frac{1}{2\pi i} \int\limits_{c-i\infty}^{c+i\infty}\!d\eta\,
    \frac{\Gamma^2(1+\eta)}{\Gamma^2(1-\eta)}\,x^{-\eta}\,
    \exp\Bigg[ - C_F\!\!\int\limits_{\alpha_s(\mu)}^{\alpha_s(\rho)}\!\frac{d\alpha}{\pi}\,
    \Ha\big(-\eta+a_\Gamma(\mu,\mu_\alpha)\big) + {\cal O}(\alpha_s^2) \Bigg] \,.
\end{equation}
Likewise, from (\ref{inverseLaplace}) and (\ref{gtdef}) we obtain
\begin{equation}
\begin{aligned}
   \sqrt{x}\,F_S^{\rm inv}(x,\mu,\rho)
   &= \frac{1}{2\pi i} \int\limits_{c-i\infty}^{c+i\infty}\!d\eta\,
    \frac{\Gamma^2(1-\eta)}{\Gamma^2(1+\eta)}\,x^\eta\,
    \exp\Bigg[ C_F\!\!\int\limits_{\alpha_s(\mu)}^{\alpha_s(\rho)}\!\frac{d\alpha}{\pi}\,
    \Ha\big(-\eta+a_\Gamma(\mu,\mu_\alpha)\big) + {\cal O}(\alpha_s^2) \Bigg] \,,
\end{aligned}
\end{equation}
where in an intermediate step we have defined the Laplace transform of the function $s(w,\mu,\rho)$ in analogy with (\ref{Stilde}). Changing the sign of the integration variable $\eta$ brings the factor in front of the exponential to the same form as in (\ref{FSdef}). 

We now use the fact that the arguments of the exponentials in both expressions can be expanded in a perturbative series in $\alpha_s(\mu)$, with coefficients that depend on the ratio $\alpha_s(\rho)/\alpha_s(\mu)$. Expanding the transfer functions at NLO in the form
\begin{equation}\label{FSexpand}
   F_S(x,\mu,\rho) = F_{S,0}(x) + \frac{C_F\alpha_s(\mu)}{\pi}\,F_{S,1}(x,r)
    + {\cal O}(\alpha_s^2) \,; \quad
   r = \frac{\alpha_s(\rho)}{\alpha_s(\mu)} \,,
\end{equation}
and similarly for $F_S^{\rm inv}(x,\mu,\rho)$, we obtain
\begin{equation}\label{ohsonice}
\begin{aligned}
   F_{S,0}(x) &= F_{S,0}^{\rm inv}(x) 
    = \frac{1}{\sqrt x}\,\MeijerG[\big]{2}{0}{0}{4}{\phantom{x}}{1,\, 1,\, 0,\, 0}{x} \,, \\[1mm]
   F_{S,1}(x,r) 
   &= - \frac{1}{\sqrt x}\,\int_0^1\!\frac{dy}{1-y}\,\frac{h(y)}{\beta_0}\,   
    \frac{r^{1+\frac{2C_F}{\beta_0} \ln y}-1}{1+\frac{2C_F}{\beta_0} \ln y}\,
    \MeijerG[\big]{2}{0}{0}{4}{\phantom{x}}{1,\, 1,\, 0,\, 0}{xy} \,, \\[-1mm]
   F_{S,1}^{\rm inv}(x,r) 
   &= \frac{1}{\sqrt x}\,\int_0^1\!\frac{dy}{1-y}\,\frac{h(y)}{\beta_0}\,  
    \frac{r^{1+\frac{2C_F}{\beta_0} \ln y}-1}{1+\frac{2C_F}{\beta_0} \ln y}\,
    \MeijerG[\Big]{2}{0}{0}{4}{\phantom{x}}{1,\, 1,\, 0,\, 0}{\frac{x}{y}} \,.
\end{aligned}
\end{equation}
To arrive at this form we have employed the leading-order approximation for $a_\Gamma(\mu,\mu_\alpha)$ shown in (\ref{eq41}) and the definition (\ref{Hdef}) relating $\Ha(a)$ to the function $h(x)$ in (\ref{hdef}). The solutions involve another example of a Meijer $G$-function. An alternative expression involving a hypergeometric function and its derivative can be derived by performing the integral in (\ref{FSdef}) using the theorem of residues. This is discussed further in Appendix~\ref{app:B}.

While the transfer functions for the soft function are given by rather complicated expressions, they nevertheless enjoy some nice properties. Combining the two relations in (\ref{Strafo}) one can derive the orthonormality condition (with $a,b>0$)
\begin{equation}\label{eq68}
   \int_0^\infty\!dx\,F_S(xa,\mu,\rho)\,F_S^{\rm inv}(xb,\mu,\rho) = \delta(a-b) \,.
\end{equation}
To prove this identity directly one uses the useful relation
\begin{equation}\label{Dirac}
   \int_0^\infty\!\frac{dx}{x}\,x^{\eta'-\eta} 
   = \int_{-\infty}^\infty\!dy\,e^{ity}
   = 2\pi\,\delta(t)
   = 2\pi i\,\delta(\eta'-\eta) \,,
\end{equation}
which holds for purely imaginary $(\eta'-\eta)=it$. We also find that powers of $x$ are very simply transformed into the diagonal space, because 
\begin{equation}\label{cuty}
   \int_0^\infty\!\frac{dx}{\sqrt{x}}\,F_S(x,\mu,\rho)\,x^b 
   = \frac{\Gamma^2(1+b)}{\Gamma^2(1-b)}
    \exp\Bigg[ - C_F\!\!\int\limits_{\alpha_s(\mu)}^{\alpha_s(\rho)}\!\frac{d\alpha}{\pi}\,
    \Ha\big(-b+a_\Gamma(\mu,\mu_\alpha)\big) + {\cal O}(\alpha_s^2) \Bigg] \,.
\end{equation}
Expanding this relation in powers of $b$ yields the transformation rules for positive integer powers of $\ln x$. 

\subsection[Convolution $T_3$ in the diagonal space]{\boldmath Convolution $T_3$ in the diagonal space}

\begin{figure}
\begin{center}
\includegraphics[width=0.6\textwidth]{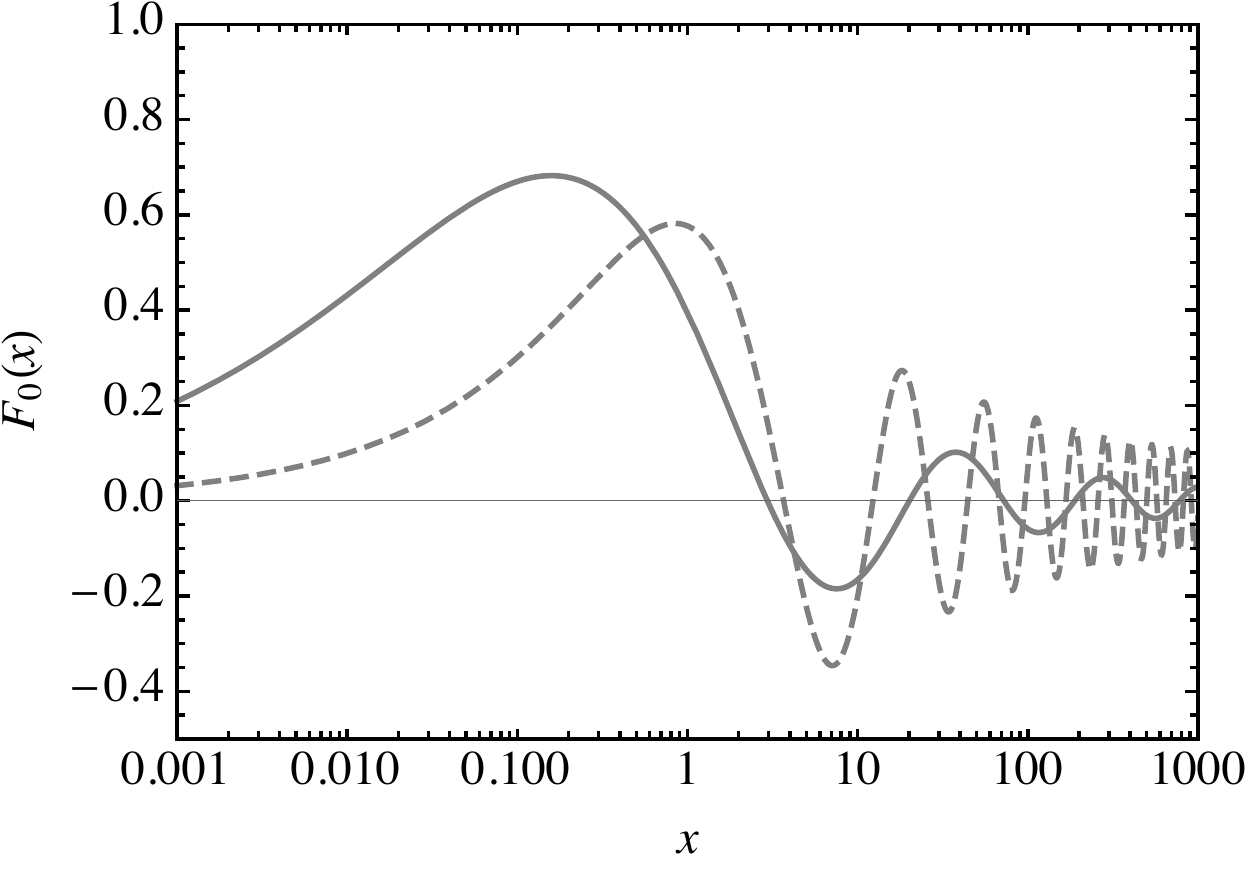}
\vspace{2mm}
\caption{\label{fig:transfer} 
Leading-order transfer functions for the soft function (solid) and the jet function (dashed).} 
\end{center}
\end{figure}

Remarkably, the factorization formula (\ref{T3ren}) retains its form in the diagonal space, 
i.e.\ 
\begin{equation}\label{T3dualtrue}
   T_3 = H_3(\mu) \int_0^\infty\!\frac{d\ell_-}{\ell_-} \int_0^\infty\!\frac{d\ell_+}{\ell_+}\,
    j(M_h\ell_-,\mu,\rho)\,j(-M_h\ell_+,\mu,\rho)\,s(\ell_+\ell_-,\mu,\rho) \,.
\end{equation}
Here $j(p^2,\mu,\rho)$ denotes the jet function in the diagonal space. To derive this relation we need the inverse transfer function for the jet function in the diagonal space, which we define as
\begin{equation}
   J(p^2,\mu) 
   = \int_0^\infty\!\frac{dx}{\sqrt{x}}\,F_J^{\rm inv}(x,\mu,\rho)\,j(p^2/x,\mu,\rho) \,.
\end{equation}
In analogy with (\ref{FSdef}) we find (drawing on expressions derived in \cite{Liu:2020ydl})
\begin{equation}\label{FJtransfertrue}
   \sqrt{x}\,F_J^{\rm inv}(x,\mu,\rho)
   = \frac{1}{2\pi i} \int\limits_{c-i\infty}^{c+i\infty}\!d\eta\,
    \frac{\Gamma(1+\eta)}{\Gamma(1-\eta)}\,x^{-\eta}\,
    \exp\Bigg[ - C_F\!\!\int\limits_{\alpha_s(\mu)}^{\alpha_s(\rho)}\!\frac{d\alpha}{2\pi}\,
    \Ha\big(-\eta+a_\Gamma(\mu,\mu_\alpha)\big) + {\cal O}(\alpha_s^2) \Bigg] \,.
\end{equation}
Expanding this function in a perturbative series as in (\ref{FSexpand}) we obtain
\begin{equation}
\begin{aligned}
   F_{J,0}^{\rm inv}(x) 
   &= {\rm J}_1(2\sqrt{x}) \,, \\
   F_{J,1}^{\rm inv}(x,r) 
   &= - \frac12\,\int_0^1\!\frac{dy}{1-y}\,\frac{h(y)}{\beta_0}\,  
    \frac{r^{1+\frac{2C_F}{\beta_0} \ln y}-1}{1+\frac{2C_F}{\beta_0} \ln y}\,
    \sqrt{y}\,\,{\rm J}_1(2\sqrt{xy}) \,.
\end{aligned}
\end{equation}
To establish that relation (\ref{T3dualtrue}) holds we need to show that
\begin{equation}\label{testcase}
   F_S(x,\mu,\rho) 
   = \int_0^\infty\!\frac{dz}{z}\,F_J^{\rm inv}(z,\mu,\rho)\,
    F_J^{\rm inv}\Big(\frac{x}{z},\mu,\rho\Big) \,.
\end{equation}
Using the representation for $F_J^{\rm inv}(x)$ given in (\ref{FJtransfertrue}) along with the identity (\ref{Dirac}) we find that the right-hand side of (\ref{testcase}) indeed leads to the expression for the soft transfer function shown in (\ref{FSdef}). In Figure~\ref{fig:transfer} we show the behavior of the leading-order transfer functions $F_{S,0}(x)$ for the soft function and $F_{J,0}(x)={\rm J}_1(2\sqrt{x})$ for the jet function. Both functions vanish in the limit $x\to 0$ and oscillate for values $x>1$. The transfer function for the jet function oscillates more rapidly, but apart from this the two functions exhibit a rather similar behavior. 

It is important that the jet functions and the soft function in (\ref{T3dualtrue}) are evaluated at the same auxiliary scale $\rho$. Only then the dependence on $\rho$ cancels out. In analogy with (\ref{dlogrhoS}), we find that the jet function in the diagonal space satisfies the different equation
\begin{equation}\label{dlogrhoJ}
   \frac{d}{d\ln\rho}\,j(p^2,\mu,\rho)
   = \bigg[ - C_F \left( \frac{\alpha_s(\rho)}{2\pi} \right)^2 \int_0^1\!\frac{dx}{1-x}\,
    h(x)\,x^{a_\Gamma(\mu,\rho)} + {\cal O}(\alpha_s^3) \bigg]\,j(x p^2,\mu,\rho) \,.
\end{equation}
It is then straightforward to check that the double convolution on the right-hand side of (\ref{T3dualtrue}) is independent of $\rho$.

\subsection{Leading-order soft function in the diagonal space}
\label{subsec:7.3}

Given the transformations in (\ref{Strafo}) we can calculate the soft function $s(w,\mu,\rho)$ in fixed-order perturbation theory from the expression for the original soft function shown in (\ref{softrenorm}). In this way we obtain $s(w,\mu_s,\rho)$ in the form of a numerical integral, where $\mu_s$ denotes the matching scale, at which a fixed-order expansion is well behaved. We can then evolve the obtained result to a different renormalization scale $\mu\ne\mu_s$ using the explicit solution (\ref{sLOsol}) of the local RG equation in the diagonal space.

\begin{figure}
\begin{center}
\includegraphics[width=0.48\textwidth]{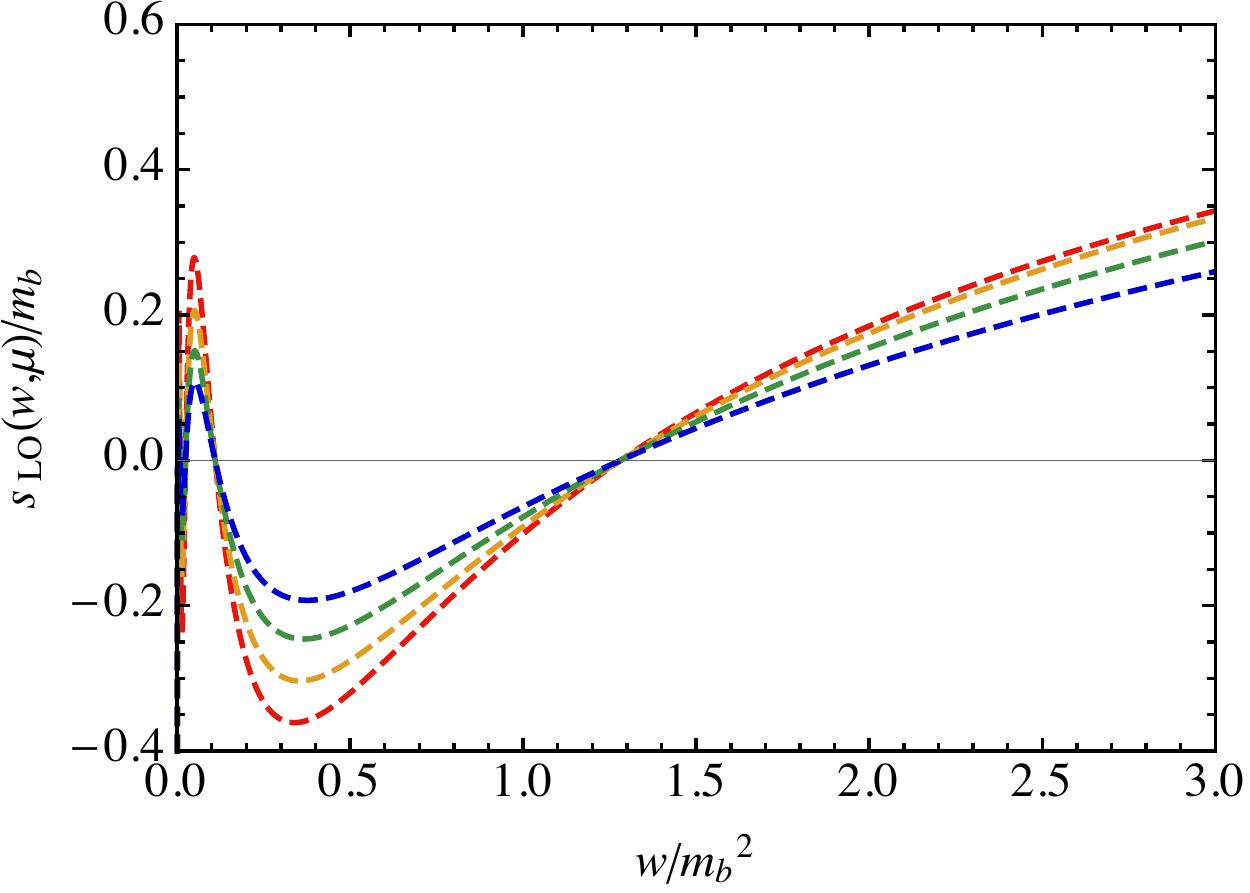}\quad
\includegraphics[width=0.48\textwidth]{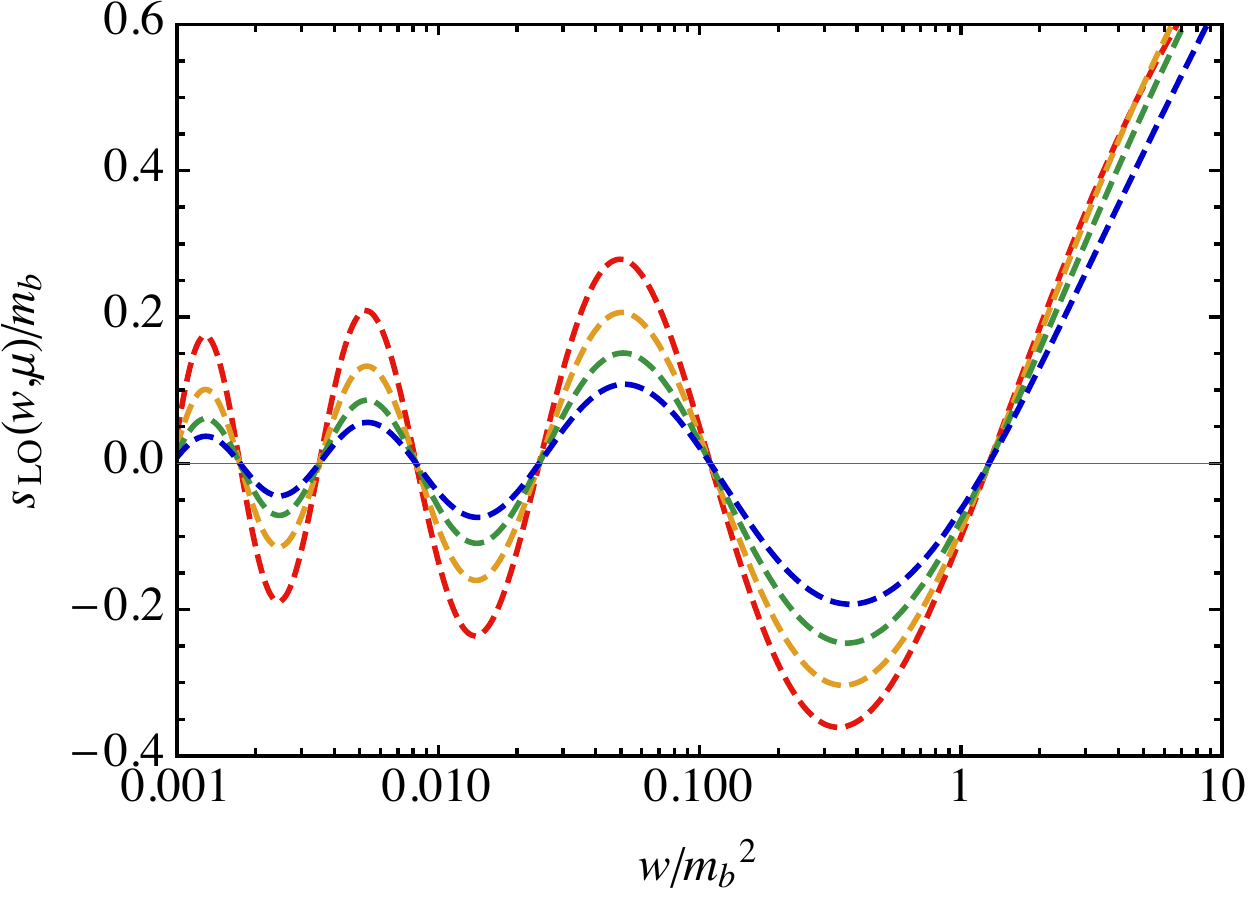}
\vspace{2mm}
\caption{\label{fig:sdualevol} 
Renormalized soft function $s_{\rm LO}(w,\mu)/m_b$ in the diagonal space at leading order in RG-improved perturbation theory for $\mu=m_b$ (red), 10\,GeV (orange), 20\,GeV (green) and 40\,GeV (blue). The plot on the right illustrates the rapid oscillations of these functions for small values of $w$.}
\end{center}
\end{figure}

At leading order in RG-improved perturbation theory it is possible to derive a simple formula for $s(w,\mu,\rho)$, which in this approximation is independent of the auxiliary scale $\rho$. Using that at the matching scale
\begin{equation}
   S(w,\mu_s) = m_b\,\theta(w-m_b^2) + {\cal O}(\alpha_s) \,,
\end{equation}
one can express the RG-evolved soft function in the diagonal space in terms of another Meijer $G$-function. We find 
\begin{equation}\label{sLO}
\begin{aligned}
   s_{\rm LO}(w,\mu) 
   &= m_b\,U_S(w;\mu,\mu_s)\,\int_0^\infty\!\frac{dx}{\sqrt{x}}\,F_{S,0}(x)\,\theta(xw-m_b^2) \\
   &= m_b\,U_S(w;\mu,\mu_s)\,
    \MeijerG[\bigg]{2}{0}{0}{4}{\phantom{x}}{0,\, 1,\, 0,\, 0}{\frac{m_b^2}{w}} \,.
\end{aligned}
\end{equation}
Note that the last form is structurally similar to the momentum-space result shown in (\ref{RGEsolLO}). 

In Figure~\ref{fig:sdualevol} we show the leading-order solution for the soft function $s_{\rm LO}(w,\mu)$ at different values of the renormalization scale $\mu$ and fixed matching scale $\mu_s=m_b=4.8$\,GeV. The behavior of the results in the region $w>m_b^2$ is qualitatively similar to the behavior seen in momentum space, see Figure~\ref{fig:Sevol}. However, for $w\ll m_b^2$ the soft function in diagonal space undergoes rapid oscillations, as shown in the right panel. These oscillations do not appear to have a physical interpretation, but they arise as a consequence of the peculiar integral transformation (\ref{Strafo}) that links functions in the diagonal space with those in momentum space. Beyond the leading order it is difficult to obtain results for the soft function in the diagonal space, because the oscillatory behavior of the transfer function makes it challenging to evaluate the slowly converging integral on the right-hand side of the first relation in (\ref{Strafo}) numerically.

\section{\boldmath RG-invariance of the convolution integral $T_3$}
\label{sec:7}

Several of the derivations performed in this paper rely on the assumption that one can pretend that the convolution integral 
\begin{equation}\label{onceagain}
   T_3 = H_3(\mu) \int_0^\infty\!\frac{d\ell_-}{\ell_-}
    \int_0^\infty\!\frac{d\ell_+}{\ell_+}\,J(M_h\ell_-,\mu)\,J(-M_h\ell_+,\mu)\,
    S(\ell_+\ell_-,\mu)
\end{equation}
is RG invariant, even though this quantity is ill defined. We will now present arguments indicating that this assumption can indeed be justified. However, we will also see that in order to properly define the quantity $T_3$ one needs to introduce a rapidity regulator, and that a {\em generic\/} rapidity regularization scheme is in conflict with RG invariance.

To see what the problem is, let us evaluate the right-hand side of (\ref{onceagain}) using the expression for the jet function obtained at leading order in RG-improved perturbation theory \cite{Liu:2020ydl}, which reads
\begin{equation}
   J_{\rm LO}(p^2,\mu) = \exp\Big[ - 2 S(\mu_j,\mu) - a_{\gamma'}(\mu_j,\mu) \Big]\,
    \frac{\Gamma\big(1-a_\Gamma(\mu_j,\mu)\big)}{\Gamma\big(1+a_\Gamma(\mu_j,\mu)\big)}
    \left(\! \frac{-p^2 e^{-2\gamma_E}}{\mu_j^2} \!\right)^{a_\Gamma(\mu_j,\mu)} ,
\end{equation}
where $\mu_j$ is an appropriate matching scale. This leads to
\begin{equation}
\begin{aligned}
   T_{3,\rm LO} &= H_3(\mu)\,\exp\Big[ - 4 S(\mu_j,\mu) - 2 a_{\gamma'}(\mu_j,\mu) \Big]\,
    \frac{\Gamma^2\big(1-a_\Gamma(\mu_j,\mu)\big)}{\Gamma^2\big(1+a_\Gamma(\mu_j,\mu)\big)} \\
   &\quad\times \int_0^\infty\!\frac{dw}{w}\,S(w,\mu)
    \left( \frac{-M_h^2\,w e^{-4\gamma_E}}{\mu_j^4} \right)^{a_\Gamma(\mu_j,\mu)} 
    \int_0^\infty\!\frac{d\ell_-}{\ell_-} \,,
\end{aligned}
\end{equation}
which is undefined. Note that RG resummation can cure the divergence of the first integral $w\to\infty$, but it does not help to regularize the divergent integral over $\ell_-$. We could have chosen to use two different matching scales $\mu_{j-}$ and $\mu_{j+}$ for the two jet functions, which would generate an extra factor $\ell_-^{\,a_\Gamma(\mu_{j-},\mu_{j+})}$ but still leaves a divergent integral. The divergence results from the fact that we are integrating over an infinite range of rapidities in the contribution $T_3$ and it requires regularization. 

\subsection{RG invariance and rapidity regularization} 
\label{subsec:7.1}

Two different rapidity regularization schemes were considered in \cite{Liu:2019oav}, and in both cases it was observed that the convolution $T_3$ by itself is not RG invariant. In that paper the regulators were applied to expression (\ref{T3}), which is written in terms of bare functions. Here we will follow an analogous approach and apply the regulators to expression (\ref{onceagain}), which is formulated in terms of renormalized functions. In the first scheme, one introduces an analytic rapidity regulator $\delta$ and an associated scale $\nu$ \cite{Becher:2010tm,Chiu:2012ir}. It has been shown in \cite{Liu:2019oav} that a consistent way to do this, which preserves the analytic properties of the $h\to\gamma\gamma$ decay amplitude, is to replace 
\begin{equation}\label{analytic}
   \int_0^\infty\!\frac{d\ell_-}{\ell_-} \int_0^\infty\!\frac{d\ell_+}{\ell_+}
   \to \int_0^\infty\!\frac{d\ell_-}{\ell_-} \int_0^\infty\!\frac{d\ell_+}{\ell_+}
    \left( \frac{M_h(\ell_+ - \ell_-)-i0}{\nu^2} \right)^{-2\delta} .
\end{equation}
In the second scheme one uses hard cutoffs on the integrals over $\ell_\pm$ in such a way that
\begin{equation}\label{cutoff}
   \int_0^\infty\!\frac{d\ell_-}{\ell_-} \int_0^\infty\!\frac{d\ell_+}{\ell_+}
   \to \lim_{\sigma\to-1}\,\int_0^{M_h}\!\frac{d\ell_-}{\ell_-} 
    \int_0^{\sigma M_h}\!\frac{d\ell_+}{\ell_+} \,.
\end{equation} 
The parameter $\sigma$ should be thought of as being a positive quantity. Only after evaluating the integrals must one take the limit $\sigma\to(-1-i0)$ by analytic continuation. 

When deriving the RG equation for the soft function based on the hypothesis of the RG invariance of $T_3$ it was important that the integrals in (\ref{onceagain}) run from 0 to $\infty$ and that the integration measures $d\ell_\pm/\ell_\pm$ are invariant under rescalings of these variables. Imposing rapidity regulators as shown above is, in general, not compatible with this derivation. Indeed, using the RG equations for the hard, jet and soft functions given in (\ref{RGEH3}), (\ref{RGE}) and (\ref{Sevol}), along with the relations (\ref{gammaS}) and (\ref{gams}), we find that after the regulators are imposed the convolution integral $T_3$ is no longer RG invariant. In the analytic regularization scheme we obtain
\begin{equation}\label{T3eq91}
\begin{aligned}
   \frac{dT_3^{\rm analytic}}{d\ln\mu} 
   &= H_3(\mu) \int_0^\infty\!dx\,K(x,\mu)\,\int_0^\infty\!\frac{d\ell_-}{\ell_-}
    \int_0^\infty\!\frac{d\ell_+}{\ell_+}\,S(\ell_+\ell_-,\mu) \\
   &\quad\times \Bigg\{ \bigg[\! \left(\! \frac{M_h(\ell_+ - x\ell_-)}{\nu^2} \!\right)^{-2\delta}\!
    - \left(\! \frac{M_h(\ell_+ - \ell_-)}{\nu^2} \!\right)^{-2\delta} \bigg]\,
    J(xM_h\ell_-,\mu)\,J(-M_h\ell_+,\mu) \\
   &\hspace{1.05cm}\mbox{}+ \bigg[\! 
    \left(\! \frac{M_h(x\ell_+ - \ell_-)}{\nu^2} \!\right)^{-2\delta}\!
    - \left(\! \frac{M_h(\ell_+ - \ell_-)}{\nu^2} \!\right)^{-2\delta} \bigg]\,
    J(M_h\ell_-,\mu)\,J(-xM_h\ell_+,\mu) \Bigg\} ,
\end{aligned}
\end{equation}
where we have dropped the ``$-i0$'' prescription for brevity. In the cutoff scheme we obtain instead
\begin{equation}\label{T3eq92}
\begin{aligned}
   \frac{dT_3^{\rm cutoff}}{d\ln\mu} 
   &= \lim_{\sigma\to -1}\,H_3(\mu) \int_0^\infty\!dx\,K(x,\mu) \\
   &\quad\times \Bigg[ \int_{M_h}^{M_h/x}\!\frac{d\ell_-}{\ell_-}
    \int_0^{\sigma M_h}\!\frac{d\ell_+}{\ell_+}\,
    J(xM_h\ell_-,\mu)\,J(-M_h\ell_+,\mu)\,S(\ell_+\ell_-,\mu) \\
   &\hspace{1.0cm}\mbox{}+ \int_0^{M_h}\!\frac{d\ell_-}{\ell_-} 
    \int_{\sigma M_h}^{\sigma M_h/x}\!\frac{d\ell_+}{\ell_+}\,
    J(M_h\ell_-,\mu)\,J(-xM_h\ell_+,\mu)\,S(\ell_+\ell_-,\mu) \Bigg] \,.
\end{aligned}
\end{equation}
In both cases the kernel $K$ is given by
\begin{equation}\label{Kdef}
   K(x,\mu) = \Gamma_{\rm cusp}(\alpha_s)\,\Gamma(1,x)
    + C_F \left( \frac{\alpha_s}{2\pi} \right)^2 \frac{\theta(1-x)}{1-x}\,h(x)
    + {\cal O}(\alpha_s^3) \,.
\end{equation}
Note that ``local'' terms in the RG equations (those with $x=1$) do not contribute in (\ref{T3eq91}) and (\ref{T3eq92}). To show that RG invariance is lost it suffices to work with the lowest-order approximations for all quantities involved, namely\footnote{We have chosen the normalization of the hard function $H_3$ differently from \cite{Liu:2019oav} in order to compensate for the different normalization of our soft function.}
\begin{equation}\label{dummies}
\begin{aligned}
   H_3(\mu) &= \frac{N_c\alpha_b}{\pi}\,\frac{y_b}{\sqrt2} + {\cal O}(\alpha_s) \,, \qquad
   J(p^2,\mu) = 1 + {\cal O}(\alpha_s) \,, \\
   S(w,\mu) &= m_b\,\theta(w-m_b^2) + {\cal O}(\alpha_s) \,.
\end{aligned}
\end{equation}
We then obtain
\begin{equation}\label{RGbreak1}
\begin{aligned}
   \frac{dT_3^{\rm analytic}}{d\ln\mu} 
   &= \frac{N_c\alpha_b}{\pi}\,\frac{y_b}{\sqrt2}\,m_b\,\frac{C_F\alpha_s}{4\pi}
    \left( \frac{-M_h^2\,m_b^2}{\nu^4} \right)^{-\delta} 
    \frac{8\Gamma^2(\delta)}{\Gamma(1+2\delta)} \left[ - H(\delta) - H(-\delta) \right]
    + {\cal O}(\alpha_s^2) \,, \\
   \frac{dT_3^{\rm cutoff}}{d\ln\mu} 
   &= \frac{N_c\alpha_b}{\pi}\,\frac{y_b}{\sqrt2}\,m_b\,\frac{C_F\alpha_s}{4\pi}\,16\zeta_3 
    + {\cal O}(\alpha_s^2) \,.
\end{aligned}
\end{equation}
In the limit $\delta\to 0$ the first expression gives the same result as the second one.

Can one find a better rapidity regularization scheme that preserves RG invariance? A third possibility, in which a cutoff $|y|<y_{\rm cut}$ is placed on the rapidity $y=\frac12\ln\frac{\ell_+}{\ell_-}$ of the soft momentum, is explored in Appendix~\ref{app:E}. It offers a small improvement on the schemes discussed above in the sense that the breaking of RG invariance occurs first at two-loop order. Yet the fundamental question posed above remains. The key to answering it is the observation that the breaking of RG invariance is linked to the non-locality of the kernel function $K(x,\mu)$, i.e.\ the fact that it does not vanish for $x\ne 1$. In Section~\ref{sec:dual} we have constructed the diagonal space, in which the evolution is strictly local in $w$ and hence $K(x,\mu)$ vanishes for $x\ne 1$. 
In this space the {\em integrand\/} of the convolution (\ref{T3dualtrue}) is RG invariant at each point in the $(\ell_+,\ell_-)$ plane, and thus RG invariance is preserved for all of our rapidity regularization schemes. 

The fact that one can find RG-invariant regularization schemes implies that the logic followed in this paper is self consistent. Assuming that $T_3$ is RG invariant is then a sensible and well-defined thing to do, and from this hypothesis we have derived the two-loop evolution equations studied in this work. Good things never come without a price, though. While imposing rapidity regulators in the diagonal space does not destroy RG invariance, it {\em does\/} interfere with the cancellation of the dependence on the auxiliary scale $\rho$ introduced in the construction of the diagonal space. The differential equations (\ref{dlogrhoS}) and (\ref{dlogrhoJ}) are then no longer sufficient to guarantee that $T_3$ as defined in (\ref{T3dualtrue}) is independent of $\rho$. This is not a serious problem, however, since $T_3$ by itself is not a physical quantity. It is only one out of three terms in a factorization theorem.

\subsection[Regularized convolution $T_3$ in the diagonal space]{\boldmath Regularized convolution $T_3$ in the diagonal space}

We have just shown that it is possible to define an RG-invariant convolution $T_3$ by applying the rapidity regularization in the diagonal space. What we have not yet demonstrated is that this convolution is well defined. At fixed order in perturbation theory the integral over $w$ diverges at infinity.\footnote{There is no divergence for $w\to 0$, because the soft function vanishes at the origin.} 
We will now show that this endpoint divergence is removed after RG improvement. It suffices to show this at leading order in RG-improved perturbation theory, because higher-order corrections do not change the functional form of the solutions in a significant way. For concreteness, we employ the analytic regularization scheme (\ref{analytic}), where the regularized convolution $T_3$ in the diagonal space takes the form
\begin{equation}
\begin{aligned}
   T_3^{\rm analytic} 
   &= H_3(\mu) \int_0^\infty\!\frac{d\ell_-}{\ell_-} \int_0^\infty\!\frac{d\ell_+}{\ell_+}\,
    j(M_h\ell_-,\mu)\,j(-M_h\ell_+,\mu)\,s(\ell_+\ell_-,\mu) \\
   &\quad\times \left( \frac{M_h(\ell_+ - \ell_-)-i0}{\nu^2} \right)^{-2\delta} .
\end{aligned}
\end{equation}
We now use the following solutions for the various component functions, valid at leading order in RG-improved perturbation theory:
\begin{equation}
\begin{aligned}
   H_{3,{\rm LO}}(\mu) 
   &= \frac{N_c\alpha_b}{\pi}\,\frac{y_b(\mu_h)}{\sqrt2}\,
    \exp\Big[ 2 S(\mu_h,\mu) - 2 a_{\gamma_q}(\mu_h,\mu) \Big] 
    \left( \frac{-M_h^2}{\mu_h^2} \right)^{-a_\Gamma(\mu_h,\mu)} \,, \\
   j_{\rm LO}(p^2,\mu)
   &= \exp\Big[ - 2 S(\mu_j,\mu) - a_{\gamma'}(\mu_j,\mu) \Big]
    \left( \frac{-p^2 e^{-2\gamma_E}}{\mu_j^2} \right)^{a_\Gamma(\mu_j,\mu)} \,, \\
   s_{\rm LO}(w,\mu)
   &= m_b\,\exp\Big[ 2 S(\mu_s,\mu) + a_{\gamma_s}(\mu_s,\mu) \Big]
    \left( \frac{w e^{-4\gamma_E}}{\mu_s^2} \right)^{-a_\Gamma(\mu_s,\mu)}\!\!
    \int_0^\infty\!\frac{dx}{\sqrt{x}}\,F_{S,0}(x)\,\theta(xw-m_b^2) \,.
\end{aligned}
\end{equation}
The hard matching scale $\mu_h$ must be chosen such that the hard function $H_3(\mu_h)$ is free of large logarithms. Natural choices are $\mu_h^2\approx M_h^2$ or $\mu_h^2\approx-M_h^2$. When combining the various exponentials in these expressions we use the identities 
\begin{equation}
\begin{aligned}
   a_\Gamma(\nu_1,\mu) - a_\Gamma(\nu_2,\mu) 
   &= a_\Gamma(\nu_1,\nu_2) \,, \\
   S(\nu_1,\mu) - S(\nu_2,\mu) 
   &= S(\nu_1,\nu_2) - a_\Gamma(\nu_2,\mu)\,\ln\frac{\nu_1}{\nu_2}
\end{aligned}
\end{equation}
for the RG functions as well as relation (\ref{gams}) to bring the answer to the explicitly $\mu$-independent form
\begin{equation}
\begin{aligned}
   T_{3,\,{\rm LO}}^{\rm analytic} 
   &= \frac{N_c\alpha_b}{\pi}\,\frac{y_b(\mu_h)}{\sqrt2}\,m_b\,
    \exp\Big[ 2 S(\mu_s,\mu_j) + 2 S(\mu_h,\mu_j)
    + a_{\gamma_s}(\mu_s,\mu_j) - 2 a_{\gamma_q}(\mu_h,\mu_j) \Big] \\
   &\quad\times \left( \frac{-M_h^2}{\mu_h^2} \right)^{-a_\Gamma(\mu_h,\mu_j)}
    \left( \frac{m_b^2}{\mu_s^2} \right)^{-a_\Gamma(\mu_s,\mu_j)}
    \frac{\Gamma^2(\delta)}{2\Gamma(2\delta)}
    \left( \frac{-M_h^2\,m_b^2}{\nu^4} \right)^{-\delta} \\
  &\quad\times e^{4\gamma_E\,a_\Gamma(\mu_s,\mu_j)} \int_0^\infty\frac{dw}{w}
   \left( \frac{w}{m_b^2} \right)^{-a_\Gamma(\mu_s,\mu_j)-\delta}
   \int_0^\infty\!\frac{dx}{\sqrt{x}}\,F_{S,0}(x)\,\theta(xw-m_b^2) \,.
\end{aligned}
\end{equation}
The integral in the last line can be performed by changing variables from $w$ to $z=xw/m_b^2$ and using the property (\ref{cuty}) of the transfer function. This gives
\begin{equation}\label{eq103}
   \int_1^\infty\frac{dz}{z}\,z^{-a-\delta}
    \int_0^\infty\!\frac{dx}{\sqrt{x}}\,F_{S,0}(x)\,x^{a+\delta}
   = \frac{\Gamma(a+\delta)\,\Gamma(1+a+\delta)}{\Gamma^2(1-a-\delta)} \,.
\end{equation}
Taking the limit $\delta\to 0$, we obtain the final result
\begin{equation}\label{shiningdiamond}
\begin{aligned}
   T_{3,\,{\rm LO}}^{\rm analytic} 
   &= \frac{N_c\alpha_b}{\pi}\,\frac{y_b(\mu_h)}{\sqrt2}\,m_b\,
    \exp\Big[ 2 S(\mu_s,\mu_j) + 2 S(\mu_h,\mu_j)
    + a_{\gamma_s}(\mu_s,\mu_j) - 2 a_{\gamma_q}(\mu_h,\mu_j) \Big] \\
   &\quad\times \left( \frac{-M_h^2}{\mu_h^2} \right)^{-a_\Gamma(\mu_h,\mu_j)}
    \left( \frac{m_b^2}{\mu_s^2} \right)^{-a_\Gamma(\mu_s,\mu_j)} 
    e^{4\gamma_E a}\,\frac{\Gamma(a)\,\Gamma(1+a)}{\Gamma^2(1-a)} \\
  &\quad\times \left[\, \frac{1}{\delta} + \ln\frac{\nu^4}{-M_h^2\,m_b^2} + 2\psi(1+a)
   + 2\psi(1-a) - \frac{1}{a} \,\right] ; \qquad a = a_\Gamma(\mu_s,\mu_j) \,.
\end{aligned}
\end{equation}
If the calculation is done in the cutoff regularization scheme (\ref{cutoff}) one finds the same result without the $1/\delta$ pole and with $\nu^2$ replaced by $(-M_h^2)$. Note that the integral over $z$ in (\ref{eq103}) converges for $a+\delta>0$ only, which for $\delta\to 0$ would require an unphysical choice $\mu_j<\mu_s$. Nevertheless, the above expression can be used to consistently define the result for any choice of scales by analytic continuation in $a$. The $1/\delta$ pole and the associated logarithm involving the scale $\nu$ result from the rapidity divergence and need to cancel against corresponding terms in other contributions to the factorization theorem for the $h\to\gamma\gamma$ decay amplitude.

To summarize this discussion, we find that the regularized convolution $T_3$ in the diagonal space is RG invariant to all orders of perturbation theory and free of endpoint divergences. This observation explains {\em a posteriori\/}  why our heuristic arguments used in Sections~\ref{sec:3} and \ref{sec:4} led to consistent results for the renormalization and scale evolution of the soft function. In the resummed expression for the convolution $T_3$ given above all large logarithms are contained in the RG functions $S(\mu_1,\mu_2)$ and $a_i(\mu_1,\mu_2)$ except for a single power of the rapidity logarithm. For natural choices of the matching scales $\mu_s\approx m_b$ and $|\mu_h|\approx M_h$ the terms in the second line are free of large logarithms. As a final comment, we emphasize that expression (\ref{shiningdiamond}) has no well-defined fixed-order expansion, because the limit $a\to 0$ does not exist. The reason is that the endpoint divergence of the original integral in (\ref{T3ren}) has been removed by resummation. When one attempts to ``undo'' the resummation by expanding the result in powers of $\alpha_s$ one encounters terms of order $1/\alpha_s^2$ and $1/\alpha_s$. These terms must cancel against similar terms contained in the other two contributions ($T_1$ and $T_2$) to the factorization theorem for the $h\to\gamma\gamma$ amplitude.

\section{Conclusions}
\label{sec:8}

In this work we have presented a detailed analysis of the renormalization and the scale evolution of the soft-quark soft function $S(w,\mu)$, defined in terms of the discontinuity of the soft quark propagator dressed by two finite-length Wilson lines connecting at one point. This function appears in the factorization formula for the $h\to\gamma\gamma$ decay amplitude induced by loops of light quarks, recently worked out by two of us \cite{Liu:2019oav}. The results we have obtained and the techniques we have developed have a more general importance in the context of understanding SCET factorization theorems at subleading order in power counting, where soft functions containing soft quarks dressed by Wilson lines become a generic feature. The relevance of our results thus extends beyond the practical purpose of studying the $h\to\gamma\gamma$ process. 

A central argument of our work has been the hypothesis that the third term $T_3$ in the $h\to\gamma\gamma$ factorization theorem, which involves a hard function and a double convolution of two radiative jet functions with the soft-quark soft function, should be RG invariant. This does not necessary have to be realized, but it is suggested by certain observations made in \cite{Liu:2019oav}. Based on this assumption we have derived the non-local renormalization factor $Z_S$ of the soft function at one-loop order in QCD, and we have shown that it successfully removes all the $1/\epsilon^n$ poles of the bare soft function. The cancellation is highly non-trivial due to the non-local structure of the counterterms and the fact that the leading-order bare soft function is not simply a constant. This observation puts our hypothesis on firmer grounds. From this result we have derived the one-loop anomalous dimension of the soft function, which closely resembles the structure of the anomalous dimension of the leading-twist $B$-meson light-cone distribution amplitude. Pushing further, we have used existing results for the two-loop anomalous dimensions of the hard and jet functions to present a conjecture for the two-loop anomalous dimension of the soft function. It would be highly desirable to test this result by a direct two-loop calculation of the soft function. 

Using our expression for the two-loop anomalous dimension we have presented an analytic closed-form solution to the non-local RG evolution equation satisfied by the soft function in momentum space. The result obtained at NLO in RG-improved perturbation theory, shown in (\ref{boundlessbeauty}), involves integrals over Meijer $G$-functions. We find that after RG evolution the discontinuous behavior of the soft function at $w=m_b^2$ seen in fixed-order perturbation theory is smoothed out. We have also studied the asymptotic behavior of the soft function for large values $w\gg m_b^2$ and emphasized the need for a dynamical choice of the soft matching scale in this context. We have then studied the RG evolution of the soft function in Laplace space and in the so-called diagonal space, where its RG evolution is local in the $w$ variable. The construction of the diagonal space in higher orders of perturbation theory requires a non-trivial extension of the dual-space formalism developed in \cite{Bell:2013tfa,Braun:2014owa}. We have explicitly constructed the transfer functions connecting the diagonal space with momentum space. Beyond the leading order in perturbation theory, functions in the diagonal space depend on an auxiliary scale $\rho$, which cancels in predictions for physical quantities. We have derived the differential equations governing the dependence on $\rho$.

Finally, we have studied the structure of the double convolution integral $T_3$ in more detail, finding that it requires rapidity regulators in order to be well defined. Using three different rapidity regularization schemes we have shown that introducing the regulators in momentum space breaks the RG invariance of $T_3$. However, RG invariance can be preserved by imposing the rapidity regulators in the diagonal space. We have presented an explicit expression for $T_3$ at leading order in RG-improved perturbation theory, in which large logarithms of the scale ratio $(-M_h^2/m_b^2)$ are resummed to all orders of perturbation theory. The resummation improves the behavior of the convolution at large momenta and tames an endpoint divergence for $w\to\infty$. 

The results obtained in this work constitute an important step toward understanding the many subtleties and complexities of SCET factorization beyond the leading order in power counting. Not only will they help to resum the large logarithms on the $h\to\gamma\gamma$ decay amplitude beyond the leading double-logarithmic approximation; more generally, we are confident that the techniques of solving the difficult non-local RG equations of soft functions developed in this work will find applications in many other factorization theorems of interest. 

\subsubsection*{Acknowledgements}
We are grateful to Lisa Zeune for many useful discussions. This research was supported by the Cluster of Excellence PRISMA$^+$\! (project ID 39083149) funded by the German Research Foundation (DFG). The research of Z.L.L.\ was also supported by the U.S.\ Department of Energy under Contract No.~DE-AC52-06NA25396, the LANL/LDRD program and within the framework of the TMD Topical Collaboration. The work of S.F.~was supported in part by the Director, Office of Science, Office of Nuclear Physics, of the U.S. Department of Energy under grant number DE-FG02-04ER41338. SF would also like to deeply thank the Mainz Institute for Theoretical Physics (MITP)for support during his leave of absence when this work was completed.

\begin{appendix}
\renewcommand{\theequation}{A.\arabic{equation}}
\setcounter{equation}{0}

\section{Anomalous dimensions and RG functions}
\label{app:A}

The exact solutions (\ref{boundlessbeauty}) and (\ref{boundlessbeauty2}) can be evaluated by expanding the anomalous dimensions and the QCD $\beta$-function as perturbative series in the strong coupling. We work consistently at NLO in RG-improved perturbation theory, keeping terms through order $\alpha_s$ in the expressions for the Sudakov exponent $S$ and the functions $a_\Gamma$ and $a_{\gamma_s}$. We define the expansion coefficients as
\begin{equation}
\begin{aligned}
   \Gamma_{\rm cusp}(\alpha_s) 
   &= \Gamma_0\,\frac{\alpha_s}{4\pi} + \Gamma_1 \left( \frac{\alpha_s}{4\pi} \right)^2
    + \Gamma_2 \left( \frac{\alpha_s}{4\pi} \right)^3 + \dots \,, \\
   \beta(\alpha_s) 
   &= -2\alpha_s \left[ \beta_0\,\frac{\alpha_s}{4\pi}
    + \beta_1 \left( \frac{\alpha_s}{4\pi} \right)^2
    + \beta_2 \left( \frac{\alpha_s}{4\pi} \right)^3 + \dots \right] ,
\end{aligned}
\end{equation}
and similarly for the anomalous dimension $\gamma_s$. Substituting these expansions in (\ref{RGfuns}) one obtains \cite{Becher:2006mr}
\begin{equation}
\begin{aligned}
   a_\Gamma(\mu_s,\mu)
   &= \frac{\Gamma_0}{2\beta_0} \left[ \ln\frac{\alpha_s(\mu)}{\alpha_s(\mu_s)}
    + \left( \frac{\Gamma_1}{\Gamma_0} - \frac{\beta_1}{\beta_0} \right)
    \frac{\alpha_s(\mu) - \alpha_s(\mu_s)}{4\pi} + {\cal O}(\alpha_s^2) \right] , \\
   S(\mu_s,\mu) 
   &= \frac{\Gamma_0}{4\beta_0^2}\,\Bigg\{
    \frac{4\pi}{\alpha_s(\mu_s)} \left( 1 - \frac{1}{r} - \ln r \right)
    + \left( \frac{\Gamma_1}{\Gamma_0} - \frac{\beta_1}{\beta_0}
    \right) (1-r+\ln r) + \frac{\beta_1}{2\beta_0} \ln^2 r \\
   &\hspace{1.6cm}\mbox{}+ \frac{\alpha_s(\mu_s)}{4\pi} \Bigg[ 
    \left( \frac{\beta_1\Gamma_1}{\beta_0\Gamma_0} - \frac{\beta_2}{\beta_0} 
    \right) (1-r+r\ln r)
    + \left( \frac{\beta_1^2}{\beta_0^2} - \frac{\beta_2}{\beta_0} \right)
    (1-r)\ln r \\
   &\hspace{3.6cm}\mbox{}- \left( \frac{\beta_1^2}{\beta_0^2} - \frac{\beta_2}{\beta_0}
    - \frac{\beta_1\Gamma_1}{\beta_0\Gamma_0} + \frac{\Gamma_2}{\Gamma_0}
    \right) \frac{(1-r)^2}{2} \Bigg] + {\cal O}(\alpha_s^2) \Bigg\} \,,
\end{aligned}
\end{equation}
where $r=\alpha_s(\mu)/\alpha_s(\mu_s)$. The function $a_{\gamma_s}(\mu_s,\mu)$ is given by an analogous expression. Whereas the two-loop anomalous dimensions and $\beta$-function are required for $a_\Gamma$ and $a_{\gamma_s}$, the expression for the Sudakov exponent $S$ also involves the three-loop coefficients $\Gamma_2$ and $\beta_2$.

We now list relevant coefficients of the anomalous dimensions and the QCD $\beta$-function, quoting all results in the $\overline{{\rm MS}}$ renormalization scheme. For the convenience of the reader, we also give numerical results for $N_c=3$ and $n_f=5$. The two-loop cusp anomalous dimension $\Gamma_{\rm cusp}$ was obtained long ago \cite{Korchemskaya:1992je}, while the three-loop coefficient was derived in \cite{Moch:2004pa}. The results are
\begin{equation}
\begin{aligned}
   \Gamma_0 &= 4 C_F = \frac{16}{3} \,, \\
   \Gamma_1 &= 4 C_F \left[ C_A \left( \frac{67}{9} - \frac{\pi^2}{3} \right) 
    - \frac{20}{9}\,T_F n_f \right] \approx 36.8436 \,, \\
   \Gamma_2 &= 4 C_F \Bigg[ C_A^2 \left( \frac{245}{6} - \frac{134\pi^2}{27}
    + \frac{11\pi^4}{45} + \frac{22}{3}\,\zeta_3 \right) 
    + C_A T_F n_f  \left( - \frac{418}{27} + \frac{40\pi^2}{27}
    - \frac{56}{3}\,\zeta_3 \right) \\
   &\hspace{1.5cm}\mbox{}+ C_F T_F n_f \left( - \frac{55}{3} + 16\zeta_3 \right) 
    - \frac{16}{27}\,T_F^2 n_f^2 \Bigg] 
    \approx 239.208 \,.
\end{aligned}
\end{equation}
The anomalous dimension $\gamma_s$ follows from (\ref{gams}) and can be determined up to two-loop order using the explicit expressions for $\gamma'$ and $\gamma_q$ given in \cite{Liu:2019oav} and \cite{Moch:2005id,Becher:2009qa}, respectively. We find
\begin{equation}
\begin{aligned}
   \gamma_{s,0} &= - 6 C_F = - 8 \,, \\
   \gamma_{s,1} &= C_F^2\,\big( - 3 + 4\pi^2 - 48\zeta_3 \big) 
    + C_F C_A \left( \frac{655}{27} - \frac{55\pi^2}{9} - 4\zeta_3 \right) 
    + C_F T_F\,n_f \left( - \frac{188}{27} + \frac{20\pi^2}{9} \right) \\
   &\approx - 151.280 \,.
\end{aligned}
\end{equation}
Finally, the expansion coefficients for the QCD $\beta$-function to three-loop order are
\begin{equation}
\begin{aligned}
   \beta_0 &= \frac{11}{3}\,C_A - \frac43\,T_F n_f 
    = \frac{23}{3} \,, \\
   \beta_1 &= \frac{34}{3}\,C_A^2 - \frac{20}{3}\,C_A T_F n_f
    - 4 C_F T_F n_f \approx 38.6667 \,, \\
   \beta_2 &= \frac{2857}{54}\,C_A^3 + \left( 2 C_F^2
    - \frac{205}{9}\,C_F C_A - \frac{1415}{27}\,C_A^2 \right) T_F n_f
    + \left( \frac{44}{9}\,C_F + \frac{158}{27}\,C_A \right) T_F^2 n_f^2 \\
   &\approx  180.907 \,. 
\end{aligned}
\end{equation}
We choose to work with $n_f=5$ active quark flavors, because we are mainly interested in RG evolution to a scale $\mu$ between the mass scales of the bottom and top quarks. 

In the analysis in Section~\ref{sec:6} we need the ratio of the running $b$-quark mass $m_b(\mu)$ defined in the $\overline{\rm MS}$ scheme and the pole mass $m_b$. At NLO in RG-improved perturbation theory this ratio is given by
\begin{equation}
   \frac{m_b(\mu)}{m_b} 
   = \left( \frac{\alpha_s(\mu)}{\alpha_s(m_b)} \right)^{\!-\frac{\gamma_{m,0}}{2\beta_0}}
    \left[ 1 - \frac{C_F\alpha_s(m_b)}{\pi} 
    - \frac{\gamma_{m,1}\beta_0 - \gamma_{m,0}\beta_1}{2\beta_0^2}\,
    \frac{\alpha_s(\mu)-\alpha_s(m_b)}{4\pi} + {\cal O}(\alpha_s^2) \right] ,
\end{equation}
where the one- and two-loop coefficients in the anomalous dimension of the quark mass are given by \cite{Tarrach:1980up}
\begin{equation}
   \gamma_{m,0} = - 6 C_F \,, \qquad
   \gamma_{m,1} = -3 C_F^2 - \frac{97}{3}\,C_F C_A + \frac{20}{3}\,C_F T_F n_f \,.
\end{equation}

\renewcommand{\theequation}{B.\arabic{equation}}
\setcounter{equation}{0}

\section{\boldmath Light-cone distribution amplitude of the $B$-meson}
\label{app:LCDA}

The RG evolution equation (\ref{Sevol}) and the associated anomalous dimension $\gamma_S$ in (\ref{gammaS}) are closely related to the corresponding equations for the leading-twist light-cone distribution amplitude $\phi_+^B(\omega)$ of the $B$ meson defined in heavy-quark effective theory  
\cite{Grozin:1996pq}. This quantity obeys the differential equation 
\begin{equation}\label{LCDA}
   \frac{d}{d\ln\mu}\,\phi_+^B(\omega,\mu) 
   = - \int_0^\infty\!d\omega'\,\gamma_+(\omega,\omega';\mu)\,\phi_+^B(\omega',\mu) \,,
\end{equation}
where \cite{Lange:2003ff,Liu:2020ydl,Braun:2019wyx}
\begin{equation}
\begin{aligned}
   \gamma_+(\omega,\omega';\mu) 
   &= - \left[ \Gamma_{\rm cusp}(\alpha_s)\,\ln\frac{\omega}{\mu} - \gamma_\eta(\alpha_s) \right] 
    \delta(\omega-\omega') - \Gamma_{\rm cusp}(\alpha_s)\,\omega\,\Gamma(\omega,\omega') \\
   &\quad\mbox{}- C_F \left( \frac{\alpha_s}{2\pi} \right)^2 
    \frac{\omega\,\theta(\omega'-\omega)}{\omega'(\omega'-\omega)}\,
    h\bigg(\frac{\omega}{\omega'}\bigg) + {\cal O}(\alpha_s^3) \,.
\end{aligned}
\end{equation}
The anomalous dimension $\gamma_S$ in (\ref{gammaS}) differs from $\gamma_+$ by the argument of the logarithm ($w/\mu^2$ versus $\omega/\mu$) and the factor 2 in front of the non-local terms. As a result of this, one finds that the solution to the evolution equation (\ref{LCDA}) can be written in terms of hypergeometric functions \cite{Lee:2005gza} rather than the more complicated Meijer $G$-functions in (\ref{RGEdual}).

\renewcommand{\theequation}{C.\arabic{equation}}
\setcounter{equation}{0}

\section{Solution in terms of hypergeometric functions}
\label{app:B}

In the derivations in Sections~\ref{sec:solu} and \ref{sec:trafo} we have written some results in terms of Meijer $G$-functions. Here we relate these functions to the more familiar hypergeometric functions. We begin with the solution to the RG equation of the soft function. For $a_\Gamma(\mu_s,\mu)<0$ it is possible to perform the contour integral in (\ref{eq42}) using the theorem of residues. The integrand contains double and single poles in the complex $\eta$-plane located at $\eta=-n$ and $\eta=n+a_\Gamma(\mu_s,\mu)$, where $n\in\mathbb{N}$. For $w>w'$ ($w<w'$), we can close the contour in the upper (lower) half plane and pick up the residues of the single poles. We assume that $a_\Gamma(\mu_s,\mu)>-1$, such that all of the poles in the second series lie in the lower half plane. The expression for the residues of the single poles involves $\Gamma$-functions and their derivatives. Comparing the answer with (\ref{eq42b}) and denoting $z=\text{min}(w,w')/\text{max}(w,w')<1$, we find that 
\begin{equation}\label{eq45}
\begin{aligned}
   \MeijerG[\Big]{2}{2}{4}{4}{-a,\, -a,\, 1-a,\, 1-a}{1,\, 1,\, 0,\, 0}{z}
   &= \frac{\Gamma^2(2+a)}{\Gamma^2(-a)}\,z\,G(z,a) \,, \\
   \MeijerG[\bigg]{2}{2}{4}{4}{-a,\, -a,\, 1-a,\, 1-a}{1,\, 1,\, 0,\, 0}{\frac{1}{z}}
   &= z^{1+a}\,\frac{\Gamma^2(2+a)}{\Gamma^2(-a)}\,G(z,a) \,,
\end{aligned}
\end{equation}
where
\begin{equation}
   G(z,a) = \frac{2a}{1+a}\,G_1(z,a) - \bigg[ \frac{2\pi}{\tan\pi a} + 4H(a) + \ln z \bigg]\,G_2(z,a)
    - 4 G_3(z,a) \,.
\end{equation}
Here $H(a)=\psi(1+a)+\gamma_E$ is the harmonic-number function, and the functions $G_i$ are given by
\begin{equation}
\begin{aligned}
   G_1(z,a) &= {}_4F_3(1\!+\!a,1\!+\!a,1\!+\!a,2\!+\!a; 2,2,2; z) \,, \\
   G_2(z,a) &= {}_4F_3(1\!+\!a,1\!+\!a,2\!+\!a,2\!+\!a; 1,2,2; z) \,, \\
   G_3(z,a) &= \left( \partial_{p_1} + \partial_{q_1} \right) 
    {}_4F_3(1\!+\!a,1\!+\!a,2\!+\!a,2\!+\!a; 1,2,2; z) \,.
\end{aligned}
\end{equation}
The derivatives in the last expression act on the indices of the hypergeometric function ${}_4F_3(p_1,p_2,p_3,p_4; q_1,q_2,q_3; z)$. The functions $G_i(z,a)$ are real-valued for $z<1$ and are singular in the limit $z\to 1^-$. Using the properties of the hypergeometric functions we have derived the precise form of this singularity shown in (\ref{singular}). 

We have also encountered the Meijer $G$-function in the calculation of the transfer function $F_S(x,\mu)$ needed for the calculation of the soft function in the diagonal space. At leading order one can ignore the exponential containing the integral over the function $\Ha$ in (\ref{Strafo}). The integrand of the $\eta$-integral then has double and single poles at values $\eta=-n$ with $n\in\mathbb{N}$. Evaluating the integral using the theorem of residues, we obtain 
\begin{equation}\label{Fmonster}
   \MeijerG[\big]{2}{0}{0}{4}{\phantom{x}}{1,\, 1,\, 0,\, 0}{x}
   = x\,\Big[ 2\,{}_0F_3(2,2,2;x)
    - \big( \ln x + 4\gamma_E + 4\partial_{q_1} \big)\,{}_0F_3(1,2,2;x) \Big] \,.   
\end{equation}

\renewcommand{\theequation}{D.\arabic{equation}}
\setcounter{equation}{0}

\section{Soft function in the dual space}
\label{app:D}

The dual soft function constructed following the approach of \cite{Bell:2013tfa,Braun:2014owa} is related to the original soft function via
\begin{equation}
   s_{\rm dual}(w,\mu) = \int_0^\infty\!\frac{dx}{x}\,S(x w,\mu)\,
    \frac{1}{2\pi i} \int\limits_{c-i\infty}^{c+i\infty}\!d\eta\,
    \frac{\Gamma^2(1+\eta)}{\Gamma^2(1-\eta)}\,x^{-\eta} 
   \equiv \int_0^\infty\!\frac{dx}{\sqrt{x}}\,F_{S,0}(x)\,S(x w,\mu) \,,
\end{equation}
where the transfer function $F_{S,0}(x)$ has been given in (\ref{ohsonice}). As a consequence of the fact that no higher-order corrections to the transfer function are included in this approach, the function $s_{\rm dual}(w,\mu)$ obeys the non-local RG evolution equation (\ref{RGEdual}). The technique we have developed 

\noindent
in Section~\ref{sec:solu} can be applied to solve this equation. In analogy with (\ref{eq42}) we obtain
\begin{equation}
\begin{aligned}
   s_{\rm dual}(w,\mu) 
   &= U_S(w;\mu,\mu_s)\,\int_0^\infty\!\frac{dw'}{w'}\,s_{\rm dual}(w',\mu_s) \\
   &\quad\times \frac{1}{2\pi i} \int\limits_{c-i\infty}^{c+i\infty}\!d\eta
    \left( \frac{w}{w'} \right)^\eta \Bigg[ 1 - \frac{C_F}{\beta_0\pi}
    \int\limits_{\alpha_s(\mu_s)}^{\alpha_s(\mu)}\!d\alpha \int_0^1\!\frac{dx}{1-x}\,h(x)\,
    x^{a_\Gamma(\mu_s,\mu_\alpha)-\eta} + {\cal O}(\alpha_s^2) \Bigg] \,,
\end{aligned}
\end{equation}
with the simplification that the $\Gamma$-functions appearing in (\ref{eq42}) are now absent. The integral over $\eta$ thus evaluates to a $\delta$-function rather than a Meijer $G$-function, yielding
\begin{equation}\label{boundlessbeauty2}
\begin{aligned}
   s_{\rm dual}(w,\mu) &= U_S(w;\mu,\mu_s) \\
   &\quad\times\!\Bigg[ s_{\rm dual}(w,\mu_s) - \frac{C_F\alpha_s(\mu_s)}{\pi}
    \int_0^1\!\frac{dx}{1-x}\,\frac{h(x)}{\beta_0}\,
    \frac{r^{1+\frac{2C_F}{\beta_0} \ln x}-1}{1+\frac{2C_F}{\beta_0} \ln x}\,
    s_{\rm dual}\Big(\frac{w}{x},\mu_s\Big) + {\cal O}(\alpha_s^2) \Bigg] .  
\end{aligned}
\end{equation}
It can readily be checked that this solution obeys the RG equation (\ref{RGEdual}) at two-loop order. This result is simpler than the corresponding relation (\ref{boundlessbeauty}) obtained in momentum space, but not as simple as the solution (\ref{sLOsol}) found in the diagonal space.

\renewcommand{\theequation}{E.\arabic{equation}}
\setcounter{equation}{0}

\section{Rapidity cutoff scheme}
\label{app:E}

In Section~\ref{subsec:7.1} we have evaluated the convolution integral $T_3$ using two different rapidity regularization schemes. As an interesting and rather natural alternative we now consider a third scheme, in which a cut $|y|<y_{\rm cut}$ is imposed directly on the rapidity $y=\frac12\ln\frac{\ell_+}{\ell_-}$ of the soft momentum. When this is done, the regularized convolution $T_3$ takes the form
\begin{equation}\label{eq95}
   T_3^{\rm rapidity} 
   = H_3(\mu) \int_0^\infty\!\frac{dw}{w} \int_{-y_{\rm cut}}^{y_{\rm cut}}\!dy\,
    J(M_h\sqrt{w}\,e^{-y},\mu)\,J(-M_h\sqrt{w}\,e^y,\mu)\,S(w,\mu) \,,
\end{equation}
and it is straightforward to show that 
\begin{equation}
\begin{aligned}
   \frac{dT_3^{\rm rapidity}}{d\ln\mu} 
   &= H_3(\mu) \int_0^\infty\!dx\,K(x,\mu)\,
    \int_0^\infty\!\frac{dw}{w}\,S(w,\mu)\,\int_{-y_{\rm cut}}^{y_{\rm cut}}\!dy \\
   &\quad\times \Big[ 2 J(M_h\sqrt{xw}\,e^{-y},\mu)\,J(-M_h\sqrt{xw}\,e^y,\mu) 
    - J(x M_h\sqrt{w}\,e^{-y},\mu)\,J(-M_h\sqrt{w}\,e^y,\mu) \\
   &\hspace{1.05cm}\mbox{}- J(M_h\sqrt{w}\,e^{-y},\mu)\,J(-x M_h\sqrt{w}\,e^y,\mu) \Big] \,.
\end{aligned}
\end{equation}
When one uses the lowest-order expression for the jet function from (\ref{dummies}) the right-hand side vanishes. However, this is no longer true when the ${\cal O}(\alpha_s)$ corrections to the jet function are taken into account. Hence, it follows that
\begin{equation}\label{RGbreak2}
   \frac{dT_3^{\rm rapidity} }{d\ln\mu} = {\cal O}(\alpha_s^2) 
\end{equation}
in this regularization scheme scheme.

\end{appendix}


\begin{thebibliography}{99}

\bibitem{Bauer:2001yt} 
  C.~W.~Bauer, D.~Pirjol and I.~W.~Stewart,
  Phys.\ Rev.\ D {\bf 65}, 054022 (2002)
  [hep-ph/0109045].

\bibitem{Bauer:2002nz} 
  C.~W.~Bauer, S.~Fleming, D.~Pirjol, I.~Z.~Rothstein and I.~W.~Stewart,
  Phys.\ Rev.\ D {\bf 66}, 014017 (2002)
  [hep-ph/0202088].
    
\bibitem{Beneke:2002ph} 
  M.~Beneke, A.~P.~Chapovsky, M.~Diehl and T.~Feldmann,
  Nucl.\ Phys.\ B {\bf 643}, 431 (2002)
  [hep-ph/0206152].

\bibitem{Becher:2014oda} 
  For a review, see: 
  T.~Becher, A.~Broggio and A.~Ferroglia,
  Lect.\ Notes Phys.\  {\bf 896}, pp.1 (2015)
  [arXiv:1410.1892 [hep-ph]].

\bibitem{Liu:2019oav}
  Z.~L.~Liu and M.~Neubert,
  JHEP \textbf{04}, 033 (2020)
  [arXiv:1912.08818 [hep-ph]].

\bibitem{Moult:2019mog} 
  I.~Moult, I.~W.~Stewart and G.~Vita,
  JHEP {\bf 1911}, 153 (2019)
  [arXiv:1905.07411 [hep-ph]].
  
\bibitem{Moult:2019uhz} 
  I.~Moult, I.~W.~Stewart, G.~Vita and H.~X.~Zhu,
  arXiv:1910.14038 [hep-ph].
  
\bibitem{Moult:2019vou} 
  I.~Moult, G.~Vita and K.~Yan,
  arXiv:1912.02188 [hep-ph].

\bibitem{Manohar:2002fd} 
  A.~V.~Manohar, T.~Mehen, D.~Pirjol and I.~W.~Stewart,
  Phys.\ Lett.\ B {\bf 539}, 59 (2002)
  [hep-ph/0204229].

\bibitem{Bell:2013tfa} 
  G.~Bell, T.~Feldmann, Y.~M.~Wang and M.~W.~Y.~Yip,
  JHEP {\bf 1311}, 191 (2013)
  [arXiv:1308.6114 [hep-ph]].

\bibitem{Braun:2014owa} 
  V.~M.~Braun and A.~N.~Manashov,
  Phys.\ Lett.\ B {\bf 731}, 316 (2014)
  [arXiv:1402.5822 [hep-ph]].

\bibitem{Tarrach:1980up}
  R.~Tarrach,
  Nucl.\ Phys.\ B \textbf{183}, 384-396 (1981).

\bibitem{Kotsky:1997rq} 
  M.~I.~Kotsky and O.~I.~Yakovlev,
  Phys.\ Lett.\ B {\bf 418}, 335 (1998)
  [hep-ph/9708485].

\bibitem{Akhoury:2001mz} 
  R.~Akhoury, H.~Wang and O.~I.~Yakovlev,
  Phys.\ Rev.\ D {\bf 64}, 113008 (2001)
  [hep-ph/0102105].

\bibitem{Penin:2014msa} 
  A.~A.~Penin,
  Phys.\ Lett.\ B {\bf 745}, 69 (2015)
  [Errata: Phys.\ Lett.\ B {\bf 751}, 596 (2015); 
  Phys.\ Lett.\ B {\bf 771}, 633 (2017)]
  [arXiv:1412.0671 [hep-ph]].

\bibitem{Liu:2017vkm}
  T.~Liu and A.~A.~Penin,
  Phys.\ Rev.\ Lett.\  {\bf 119}, no. 26, 262001 (2017)
  [arXiv:1709.01092 [hep-ph]].
  
\bibitem{Liu:2018czl} 
  T.~Liu and A.~Penin,
  JHEP {\bf 1811}, 158 (2018)
  [arXiv:1809.04950 [hep-ph]].

\bibitem{Becher:2006mr} 
  T.~Becher, M.~Neubert and B.~D.~Pecjak,
  JHEP {\bf 0701}, 076 (2007)
  [hep-ph/0607228].

\bibitem{Bosch:2003fc} 
  S.~W.~Bosch, R.~J.~Hill, B.~O.~Lange and M.~Neubert,
  Phys.\ Rev.\ D {\bf 67}, 094014 (2003)
  [hep-ph/0301123].

\bibitem{Liu:2020ydl} 
  Z.~L.~Liu and M.~Neubert,
  arXiv:2003.03393 [hep-ph].
  
\bibitem{Becher:2009qa} 
  T.~Becher and M.~Neubert,
  JHEP {\bf 0906}, 081 (2009)
  [Erratum: JHEP {\bf 1311}, 024 (2013)]
  [arXiv:0903.1126 [hep-ph]].

\bibitem{Korchemskaya:1992je} 
  I.~A.~Korchemskaya and G.~P.~Korchemsky,
  Phys.\ Lett.\ B {\bf 287}, 169 (1992).

\bibitem{Braun:2019wyx} 
  V.~M.~Braun, Y.~Ji and A.~N.~Manashov,
  Phys.\ Rev.\ D {\bf 100}, no. 1, 014023 (2019)
  [arXiv:1905.04498 [hep-ph]].

\bibitem{Grozin:1996pq} 
  A.~G.~Grozin and M.~Neubert,
  Phys.\ Rev.\ D {\bf 55}, 272 (1997)
  [hep-ph/9607366].

\bibitem{Lange:2003ff} 
  B.~O.~Lange and M.~Neubert,
  Phys.\ Rev.\ Lett.\  {\bf 91}, 102001 (2003)
  [hep-ph/0303082].

\bibitem{Lee:2005gza} 
  S.~J.~Lee and M.~Neubert,
  Phys.\ Rev.\ D {\bf 72}, 094028 (2005)
  [hep-ph/0509350].
  
\bibitem{Gfun2}
  A.~M.~Mathai and R.~K.~Saxena, 
  {\em Generalized Hypergeometric Functions with Applications in Statistics and Physical Sciences\/} 
  (Springer, Berlin, 1973).

\bibitem{Gfun3}
  R.~Beals and J.~Szmigielski, 
  Notices of the American Mathematical Society {\bf 872}, vol.~60, no.~7 (2013)
  [https://www.ams.org/notices/201307/rnoti-p866.pdf].

\bibitem{Becher:2007ty}
  T.~Becher, M.~Neubert and G.~Xu,
  JHEP \textbf{07}, 030 (2008)
  [arXiv:0710.0680 [hep-ph]].

\bibitem{Neubert:2005nt}
  M.~Neubert,
  Phys. Rev. D \textbf{72}, 074025 (2005)
  [arXiv:hep-ph/0506245 [hep-ph]].

\bibitem{Becher:2006nr}
  T.~Becher and M.~Neubert,
  Phys. Rev. Lett. \textbf{97}, 082001 (2006)
[arXiv:hep-ph/0605050 [hep-ph]].

\bibitem{Becher:2010tm} 
  T.~Becher and M.~Neubert,
  Eur.\ Phys.\ J.\ C {\bf 71}, 1665 (2011)
  [arXiv:1007.4005 [hep-ph]].

\bibitem{Chiu:2012ir}
  J.~Chiu, A.~Jain, D.~Neill and I.~Z.~Rothstein,
  JHEP \textbf{05}, 084 (2012)
  [arXiv:1202.0814 [hep-ph]].

\bibitem{Moch:2004pa} 
  S.~Moch, J.~A.~M.~Vermaseren and A.~Vogt,
  Nucl.\ Phys.\ B {\bf 688}, 101 (2004)
  [hep-ph/0403192].

\bibitem{Moch:2005id} 
  S.~Moch, J.~A.~M.~Vermaseren and A.~Vogt,
  JHEP {\bf 0508}, 049 (2005)
  [hep-ph/0507039].
  
\end{thebibliography}
\end{document}